\documentclass[conference]{IEEEtran}

\usepackage{cite}
\ifCLASSINFOpdf
   \usepackage[pdftex]{graphicx}
\else
\fi
\usepackage{amsmath,amssymb,amsthm}
\usepackage{steinmetz}
\usepackage{nicefrac}
\usepackage{soul}
\usepackage{multirow}

\usepackage{tikz}
\usepackage{pgfplots}
\usetikzlibrary{arrows}
\usepgfplotslibrary{units}

\usepackage{array}
\ifCLASSOPTIONcompsoc
  \usepackage[caption=false,font=normalsize,labelfont=sf,textfont=sf]{subfig}
\else
  \usepackage[caption=false,font=footnotesize]{subfig}
\fi

\IEEEoverridecommandlockouts

\newcommand{\fixme}[2]{\ifx&#2&{\color{red}#1}\else{\color{red}FIXME\{}#1{\color{red}\}}\footnote{{\color{red}#2}}\PackageWarning{Fixme}{#1: #2}\fi}
\usepackage{booktabs}
\usepackage{color}

\definecolor{Set1-7-1}{RGB}{228,26,28}
\definecolor{Set1-7-2}{RGB}{55,126,184}
\definecolor{Set1-7-3}{RGB}{77,175,74}
\definecolor{Set1-7-4}{RGB}{152,78,163}
\definecolor{Set1-7-5}{RGB}{255,127,0}
\definecolor{Set1-7-6}{RGB}{166,86,40}
\definecolor{Set1-7-7}{RGB}{0,0,0}

\newcommand{\figurewidth}{0.9}	
\newcommand{\figureheight}{0.72}	

\begin{document}

\bstctlcite{IEEEexample:BSTcontrol}

\title{On the Advantage of Coherent LoRa Detection\\in the Presence of Interference}

\author{\IEEEauthorblockN{Orion Afisiadis,\IEEEauthorrefmark{1} Sitian Li,\IEEEauthorrefmark{1} Andreas Burg\IEEEauthorrefmark{1}, and Alexios Balatsoukas-Stimming,\IEEEauthorrefmark{2}}
	\IEEEauthorblockA{\IEEEauthorrefmark{1}Telecommunication Circuits Laboratory, \'{E}cole polytechnique f\'{e}d\'{e}rale de Lausanne, Switzerland\\
		\IEEEauthorrefmark{2}Department of Electrical Engineering, Eindhoven University of Technology, The Netherlands}\\%
	Email: orion.afisiadis@epfl.ch
}

\maketitle

\begin{abstract}
It has been shown that the coherent detection of LoRa signals only provides marginal gains of around $0.7$~dB on the additive white Gaussian noise (AWGN) channel. However, ALOHA-based massive Internet of Things systems, including LoRa, often operate in the interference-limited regime. Therefore, in this work, we examine the performance of the LoRa modulation with coherent detection in the presence of interference from another LoRa user with the same spreading factor. We derive rigorous symbol- and frame error rate expressions as well as bounds and approximations for evaluating the error rates. The error rates predicted by these approximations are compared against error rates found by Monte Carlo simulations and shown to be very accurate. We also compare the performance of LoRa with coherent and non-coherent receivers and we show that the coherent detection of LoRa is significantly more beneficial in interference scenarios than in the presence of only AWGN. For example, we show that coherent detection leads to a 2.5~dB gain over the standard non-coherent detection for a signal-to-interference ratio (SIR) of 3~dB and up to a 10~dB gain for an SIR of 0~dB.
\end{abstract}

\IEEEpeerreviewmaketitle

\section{Introduction} \label{sec:intro}
LoRaWAN is a proprietary standard, which is currently one of the most popular non-cellular communications solutions for the Internet of Things~\cite{Raza2017,Haxhibeqiri2018,Boulogeorgos2016}. The physical layer of LoRaWAN, which is simply called LoRa~\cite{Seller2016}, uses symbols which are frequency chirps that span the entire allocated bandwidth~\cite{Vangelista2017}. The spreading gain of LoRa is determined by the spreading factor (SF), which can be adjusted to trade off data rate and transmission time for range and robustness. For energy efficiency reasons, LoRa uses a non-slotted ALOHA-based MAC protocol. Unfortunately, the long transmission times of LoRa packets along with this ALOHA-based protocol result in a large number of packet collisions~\cite{Georgiou2017,Bor2016,BenTemim2020}. This issue is aggravated as the number of LoRa devices increases in the future, putting the scalability of LoRa networks at risk as they become interference-limited.

We call this type of LoRa packet collisions \emph{same-technology interference}. On the contrary, interference coming from other technologies in the industrial, scientific and medical (ISM) band is called \emph{cross-technology interference}~\cite{Orfanidis2017,Marquez2020}. Same-technology interference can be divided in two main types: the first type is interference from other LoRa nodes which use different spreading factors, which is called \emph{inter-SF interference}. The second, and most severe type of same-technology interference comes from LoRa nodes transmitting with the same spreading factor, and is called \emph{same-SF interference}. The impact of same-technology interference of both types has received significant attention in the literature~\cite{Bor2016,Voigt2016,Georgiou2017,Haxhibeqiri2017b,Ferrari2017,Fernandes2019,Croce2017,Croce2018,Goursaud2015,Mikhaylov2017,Feltrin2018,Elshabrawy2018b,Elshabrawy2019,Mahmood2019,Waret2019,Croce2020,Georgiou2020}.

The scalability of LoRaWAN networks is evaluated either through mathematical models or using system-level simulations. In particular, in system-level simulators it is possible to tune many important parameters of the network, such as the number of nodes per SF, the total number of nodes in the network, the amount of transmitted data, the duty cycle of the nodes, the transmission power, etc. Such system-level simulation environments are therefore essential tools to provide performance results for different network configurations, including large and heavily-loaded networks. Many system-level LoRaWAN simulators have been proposed in the literature during the last years~\cite{Bor2016,Pop2017,Croce2017,Abdelfadeel2020,Callebaut2019,Abeele2017,Reynders2018,Reynders2018b,Magrin2020,To2018,Kouvelas2018,Slabicki2018,Yousuf2018,Centenaro2017,Finnegan2020}. All of these simulators rely on models for the performance characteristics (e.g., frame error rates) of the underlying PHY layer. Today, most of these models are highly simplified, especially when it comes to considering interference. To alleviate this issue, it is essential to provide realistic and detailed PHY performance models, especially when considering new types of modulations or receivers. Detailed PHY performance models are essential tools in a combined PHY/MAC layer approach to realistically evaluate the scalability of LoRaWAN~\cite{Furtado2020}.

The number of works that provide rigorous probabilistic models for the performance of LoRa PHY is still relatively small, but growing. The performance of LoRa under additive white Gaussian noise (AWGN) and various fading scenarios has been studied in the literature, both theoretically~\cite{Reynders2016,Reynders2016b,Elshabrawy2018,FerreiraDias_2019,Baruffa2019,Courjault2019,Afisiadis2019c,Baruffa2020, Guo2020} and through experimental measurements~\cite{Callebaut2019,Xu2020}. Moreover, the error rate of a LoRa receiver under interference from another LoRa user has been thoroughly examined in~\cite{Elshabrawy2018b} and~\cite{Afisiadis2020}. However, to date the error rate of LoRa has mostly been studied for non-coherent detection. Specifically, the performance of coherent LoRa receivers has, to the best of our knowledge, been examined in~\cite{Reynders2016b,Elshabrawy2019b,Marquet2019,Baruffa2019} for the AWGN scenario, and recently in~\cite{Marquet2020} for Rayleigh fading. Very recently, new error rate approximations have been given both for coherent and non-coherent LoRa in~\cite{Baruffa2020}. For AWGN channels, coherent LoRa receivers have been shown to yield only a relatively insignificant performance improvement of approximately $0.7$~dB over the non-coherent receiver~\cite{Elshabrawy2019b}. In the same work, the authors also investigated additional orthogonal dimensions for LoRa signaling to improve the throughput of LoRa. Since interference scenarios in LoRa networks are very common, and will become even more common in large-scale LoRa deployments in the near future, the analysis of the performance of coherent LoRa receivers under interference is an important topic that is currently missing in the literature.

\subsubsection*{Contributions}
In this work, we propose coherent detection as a simple method for LoRa receivers to significantly improve their resilience against same-SF interference. To demonstrate the potential, we model the performance of coherent LoRa receivers under same-SF interference from another LoRa user and we derive an expression for the symbol error rate (SER). Since this expression is computationally intensive to evaluate in practice, we also derive a corresponding lower bound and an approximation that are based on Q-functions. Furthermore, we extend our analysis to the frame error rate (FER), which is more useful for LoRa system-level simulators. We show that, interestingly, the performance improvement provided by coherent LoRa receivers can potentially be much larger in the case of same-SF interference than the $0.7$~dB improvement shown in the literature for the AWGN case. For example, for a signal-to-interference ratio (SIR) of $0$~dB the improvement of the coherent over the non-coherent detection is around $10$~dB. We finally note that in this work we provide a thorough interference model, which can be considered as an extension of our interference model in~\cite{Afisiadis2020}. The new extended interference model avoids oversimplifications on the discrete-time baseband equivalent representation of the interfering signal. Furthermore, the interference model includes effects such as the carrier frequency offset (CFO) of the interferer, and provides a more realistic modeling not only for the proposed coherent detection of LoRa, but also for the existing non-coherent detection.

\subsubsection*{Outline}
The remainder of this paper is organized as follows. In Section~\ref{sec:system_model}, we provide a description of the LoRa modulation and the standard non-coherent detection. Moreover, we discuss coherent LoRa detection for interference scenarios and we summarize the analysis in the existing literature regarding the error rate of coherent LoRa detection under AWGN. In Section~\ref{sec:SER_interf}, we extend the analysis to derive the coherent LoRa symbol error rate expression in the presence of interference. In Section~\ref{sec:bound_approx}, we derive lower bounds and an approximation for the coherent LoRa symbol error rate in the presence of interference based on easy-to-evaluate Q-functions. In Section~\ref{sec:FER_interf}, we derive the frame error rate expression which is of great practical interest for system-level simulators. Finally, Section~\ref{sec:results} contains numerical symbol and frame error rate results and Section~\ref{sec:conclusion} concludes this paper.

\section{Coherent LoRa Detection} \label{sec:system_model}
LoRa is a spread-spectrum modulation with typical values for bandwidth $B \in \{ 125,250,500\}$~kHz. Each LoRa symbol consists of $N = 2^\text{SF}$ chips and carries SF bits of information, where $\text{SF} \in \{7, \dots, 12\}$. In the discrete-time baseband-equivalent representation, the bandwidth $B$ is split into $N$ frequency steps. The first chip of a symbol $s \in \mathcal{S}$, where $ \mathcal{S} = \left\{0,\hdots,N{-}1\right\} $, begins at a baseband frequency of $(\frac{s B}{N} - \frac{B}{2})$. At every chip of the symbol, the frequency increases by $\frac{B}{N}$, until the Nyquist frequency $\frac{B}{2}$ is reached. When this happens, i.e., at chip $n_{\text{fold}} = N-s$, there is a frequency fold to $-\frac{B}{2}$. 

As explained in~\cite{Elshabrawy2019b}, the general discrete-time baseband-equivalent description of a LoRa symbol $s$ can be written in two forms: in the first form~\cite{Elshabrawy2019c}, the phase of the first chip of a LoRa symbol is symbol-dependent. This form is therefore not directly suited for coherent demodulation. In the second form, as shown in~\cite{Ghanaatian2019,Elshabrawy2019b,Chiani2019,Afisiadis2020}, all LoRa symbols start with the same phase. Therefore, only the second form has inter-symbol phase continuity, which has been described in the LoRa patent~\cite{Seller2016} as a desirable property. 

We follow our notation from~\cite{Ghanaatian2019,Afisiadis2020} and write the discrete-time baseband-equivalent description of a LoRa symbol with inter-symbol phase continuity, as
\begin{align} \label{eq:LoRa_symbol_twoEqs}
x_s[n] & =
\begin{cases}
e^{j2\pi\left( \frac{1}{2N} \left(\frac{B}{f_s}\right)^2n^2 + \left(\frac{s}{N} - \frac{1}{2}\right)\left(\frac{B}{f_s}\right)n \right)}, &  n \in \mathcal{S}_1,\\
e^{j2\pi\left( \frac{1}{2N} \left(\frac{B}{f_s}\right)^2n^2 + \left(\frac{s}{N} - \frac{3}{2}\right)\left(\frac{B}{f_s}\right)n \right)}, &  n \in \mathcal{S}_2,
\end{cases}
\end{align}
where $\mathcal{S}_1 = \{0,..., n_{\text{fold}}-1\}$ and $\mathcal{S}_2 = \{n_{\text{fold}},...,N-1\}$. In the case where the sampling frequency $f_s$ is equal to $B$, the discrete-time baseband-equivalent description of a LoRa symbol $s$ can be simplified to\footnote{As explained  in~\cite{Chiani2019}, the continuous-time LoRa chirp occupies a slightly larger bandwidth than B. As such, using a sampling rate $f_s = B$ in a real transmission introduces some distortion effects due to aliasing. However, as explained in~\cite{Chiani2019}, since $N$ is large in LoRa, most of the signal power lies inside B, and therefore,~\eqref{eq:LoRa_symbol} is a good discrete-time baseband-equivalent description of the LoRa signal.}
\begin{align} \label{eq:LoRa_symbol}
x_s[n] & = e^{j2\pi \left(\frac{n^2}{2N}  + \left(\frac{s}{N} - \frac{1}{2}\right)n \right)}, \;\; n \in \mathcal{S}.
\end{align}
When the transmission takes place over an AWGN channel with a given complex-valued channel gain $h \in \mathbb{C}$, the received LoRa symbol is given by
\begin{align}
y[n] & = hx_s[n] + z[n], \;\; n \in \mathcal{S}, \label{eq:lora_rx}
\end{align}
where $z[n] \sim \mathcal{CN}(0,\sigma^2)$ is complex AWGN with variance $\sigma^{2} = \frac{N_0}{2N}$ and $N_0$ is the single-sided noise power spectral density. The channel has a magnitude and a phase, i.e., $h = |h|e^{j\phi}$, where we assume $|h| = 1$ without loss of generality, so that the signal-to-noise ratio (SNR) is $\frac{1}{N_0}$.

On the receiver side, the first task is called the \emph{dechirping}, which multiplies the received signal by the complex conjugate of a reference signal $x_{\text{ref}}$. A common choice for the reference signal is the LoRa symbol for $s=0$
\begin{align}
x_{\text{ref}}[n] = e^{j2\pi \left(\frac{n^2}{2N} - \frac{n}{2} \right)}, \;\; n \in \mathcal{S}. \label{eq:Ref_symbol}
\end{align}
After dechirping, the discrete Fourier transform (DFT) of the signal is computed to obtain $\mathbf{Y} = \text{DFT}\left(\mathbf{y} \odot \mathbf{x}_{\text{ref}}^{*}\right)$, where $\odot$ denotes the Hadamard product, $\mathbf{y} = \begin{bmatrix} y[0] & \hdots & y[N-1] \end{bmatrix}$, and $\mathbf{x}_{\text{ref}} = \begin{bmatrix} x_{\text{ref}}[0] & \hdots & x_{\text{ref}}[N-1] \end{bmatrix}$. Non-coherent LoRa detection is performed by simply selecting the bin index with the largest magnitude
\begin{align}
\hat{s} = \arg\max_{k\in \mathcal{S} } \left( |Y_k| \right), \label{eq:nonCoherentDemod}
\end{align}
where $Y_k$ denotes the $k$-th element of $\mathbf{Y}$.

\subsection{Coherent LoRa Detection: How and Why?} \label{sec:WhyCoherent}

Coherent LoRa detection necessitates a phase rotation of the received signal to compensate for the channel phase rotation. The phase shift $\phi$ introduced by the transmission channel can be estimated after initial synchronization using the preamble LoRa symbols, according to
\begin{align}
\hat{\phi} = \arg \left( \sum_{i=1}^{N_{pr}} Y_{0}^{(i)}\right), \label{eq:phi_est}
\end{align}
where $N_{pr}$ is the number of preamble symbols in the LoRa packet, and $Y_{0}^{(i)}$ is the first frequency bin of the $i$-th preamble symbol.

After the compensation of the phase rotation according to $\mathbf{U} = \mathbf{Y} e^{-j\hat{\phi}} $, coherent detection is performed by selecting the bin index with the maximum projection onto the real axis~\cite{Elshabrawy2019b,Marquet2019,Baruffa2019}
\begin{align}
\hat{s} = \arg\max_{k\in \mathcal{S} } \left( \Re(U_k) \right). \label{eq:coherentDemod}
\end{align}

\begin{figure}[tb]
	\centering
	\subfloat[AWGN]{%
		\includegraphics[width=0.29\textwidth]{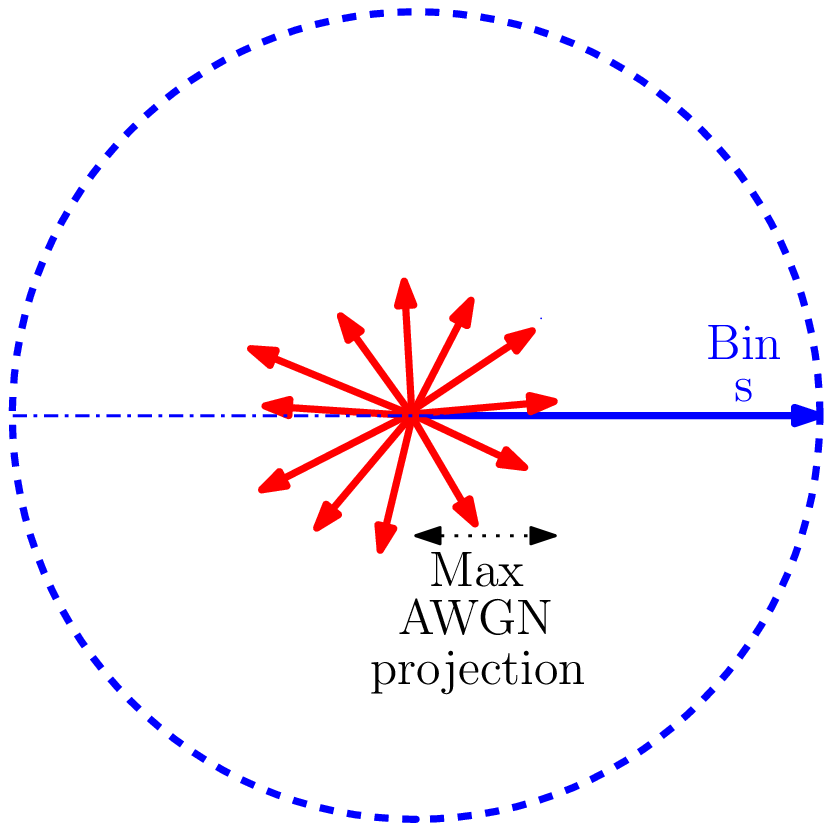}%
		\label{fig:bins_AWGN_coh}
	}\quad
	\subfloat[Same-SF interference]{%
		\includegraphics[width=0.29\textwidth]{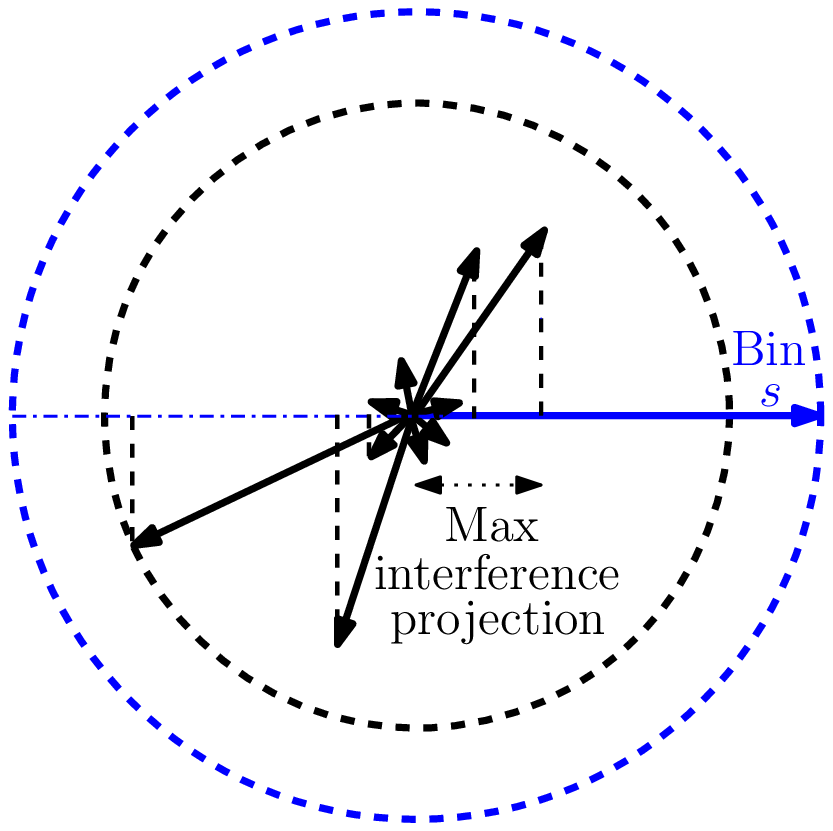}
		\label{fig:bins_coh}
	}
	\caption{Illustration of the complex plane representation of the received dechirped signal in frequency domain.}
	\label{fig:bins_coh_both}
\end{figure}

Fig.~\ref{fig:bins_coh_both} illustrates the impact of coherent detection for AWGN and with interference, by showing the elements of $\mathbf{U}$ superimposed as vectors in the complex plane. While the desired signal is aligned with the real axis, signal bins containing noise or interference are circularly symmetric. Unfortunately, for AWGN (see Fig.~\ref{fig:bins_AWGN_coh}), the  large number of noise signal components renders it likely that at least one is aligned well with the signal, a fact that limits the gain from coherent detection.
However, in the interference-limited scenario (Fig.~\ref{fig:bins_coh}), there is only a small number of interference bins that have a significant magnitude and the risk of one of these few aligning with the real axis (which increases the likelihood of an error event) is low.

\subsection{Coherent LoRa Symbol Error Rate Under AWGN}

\begin{figure*}[t]
	\begin{align} \label{eq:formula_SER_coh_only_noise_Karagiannidis}
	P(\hat{s} {\neq} s)  
	& = \frac{1}{\sigma\sqrt{2\pi}} \sum_{q{=}1}^{N{-}1}({-}1)^{q{+}1}\binom{N{-}1}{q} \int_{y{=}0}^{{+} \infty} \left(Q\left(\frac{y}{\sigma}\right)\right)^q e^{{-}\frac{\left(y{-}N\right)^2}{2\sigma^{2}}} {dy}
	\end{align}
	\hrule
\end{figure*}

Before proceeding to analyze the SER of coherent detection with interference, it useful to first recapitulate the performance of coherent detection under AWGN.

In an ideal noiseless receiver and under the assumption of perfect sample synchronization, the DFT of the dechirped signal $\mathbf{Y}$ results in a single frequency bin which contains all the signal energy (i.e., a bin that has magnitude $N$) and all remaining $N-1$ bins have zero energy. With AWGN present, the distribution of the frequency bin values $Y_k$ for $k \in \mathcal{S}$ is
\begin{align}\label{eq:distribution_Yk'}
Y_k \sim
\begin{cases}
\mathcal{CN}\left(0,2 \sigma^{2} \right), & k \in \mathcal{S} / s, \\
\mathcal{CN}\left(N\cos \phi + jN\sin \phi,2\sigma^{2} \right), & k = s. \\
\end{cases}
\end{align}
where $s$ is the transmitted symbol~\cite{Afisiadis2020}. 
The channel phase shift $\phi$ is fixed for each transmission, but is generally uniformly distributed in $[0,2\pi)$.

After the compensation for the channel phase by the coherent receiver, the corresponding demodulation metric follows a normal distribution with zero mean for $k \in \mathcal{S} / s$ and a normal distribution with mean $N$ for $k=s$
\begin{align}\label{eq:distribution_ReYk'}
\Re(U_k) \sim
\begin{cases}
\mathcal{N}\left(0,\sigma^{2} \right), & k \in \mathcal{S} / s, \\
\mathcal{N}\left(N,\sigma^{2} \right), & k = s. \\
\end{cases}
\end{align}

A symbol error occurs if and only if any of the $\Re(U_k)$ values for $k \in \mathcal{S}/s$ exceeds the value of $\Re(U_s)$, or, equivalently, if and only if $\Re(U_{\max}) > \Re(U_s)$, where $\Re(U_{\max}) = \max _{k \in \mathcal{S}/s}\Re(U_k)$. The probability density function (PDF) of bin $s$ is $f_{\Re(U_{s})} (y) = \mathcal{N}\left(N,\sigma^{2} \right)$ and the cumulative distribution function (CDF) of the maximum projection of the remaining bins on the real axis, $\Re(U_{\max})$, is 
\begin{align} \label{eq:max_order_cdf}
F_{\Re(U_{\max})} (y) = \left(1-Q\left(\frac{y}{\sigma}\right)\right)^{N{-}1},
\end{align}
where $Q(\cdot)$ denotes the Q-function. Therefore, the probability of symbol error for a given transmitted symbol $s$ is
{\small
\begin{align} \label{eq:formula_SER_coh_only_noise_finalform}
P(\hat{s} \neq s|s)  
& {=} \frac{1}{\sigma\sqrt{2\pi}} \int_{y{=}0}^{{+} \infty} \left(1{-}\left(1{-}Q\left(\frac{y}{\sigma}\right)\right)^{N{-}1}\right) e^{{-}\frac{\left(y{-}N\right)^2}{2\sigma^{2}}} {dy}. 
\end{align}
}
The SER for all symbols $s$ is identical, therefore~\eqref{eq:formula_SER_coh_only_noise_finalform} is also equal to the expected SER $P(\hat{s} \neq s)$.

The SER in~\eqref{eq:formula_SER_coh_only_noise_finalform} is equivalent to~\cite[Eq. (17)]{Baruffa2019}, where the authors explain how to evaluate it numerically without suffering from numerical problems. The authors of~\cite{Elshabrawy2019b} give a low-complexity approximation for the coherent LoRa SER under AWGN. The approximation from~\cite{Elshabrawy2019b}, which we write below using our notation, was derived using curve fitting
\begin{align}\label{eq:SER_AWGN_approx}
P(\hat{s} \neq s)  & \approx Q\left(\frac{1-\sqrt{\sigma^2}\left(1.161 + 0.2074\cdot\text{SF}\right)}{\sqrt{\sigma^2 + \sigma^2\left(0.2775-0.0153\cdot\text{SF}\right)}}\right).
\end{align}
The approximation in~\eqref{eq:SER_AWGN_approx} can be evaluated with very low complexity and is very accurate.

As an alternative to the empirical fit in~\eqref{eq:SER_AWGN_approx}, we  note here that using Newton's binomial identity, we have
\begin{align} \label{eq:NewtonsBinomial}
1{-}\left(1{-}Q\left(\frac{y}{\sigma}\right)\right)^{N{-}1}
& {=} \sum_{q{=}1}^{N{-}1}({-}1)^{q{+}1}\binom{N{-}1}{q} \left(Q\left(\frac{y}{\sigma}\right)\right)^q,
\end{align}
and thus we can write ~\eqref{eq:formula_SER_coh_only_noise_finalform} as~\eqref{eq:formula_SER_coh_only_noise_Karagiannidis}. The SER written in the form of~\eqref{eq:formula_SER_coh_only_noise_Karagiannidis} contains integer powers of the Q-function, and thus it can be evaluated using the simple and tight approximation for the integer powers of the Q-function of~\cite{Karagiannidis2007}.

\section{Coherent LoRa Symbol Error Rate \\Under Same-SF Interference} \label{sec:SER_interf}

In this section, we extend the signal model of Section~\ref{sec:system_model} for the same-SF LoRa interference scenario. We then derive the SER expression of LoRa with coherent detection under same-SF interference.

\subsection{Interference Signal Model}
\begin{figure}
	\centering
	\includegraphics[width=0.45\textwidth]{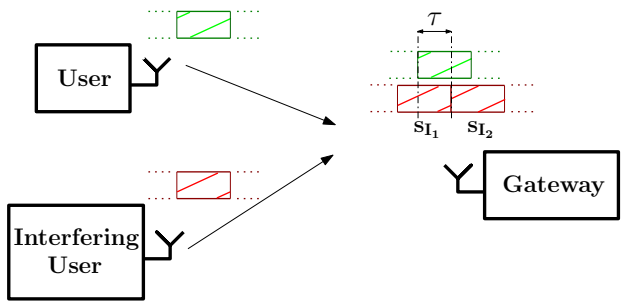}
	\caption{Illustration of LoRa uplink transmission with one interfering user having an arbitrary $\tau$~\cite{Afisiadis2020}.}
	\label{fig:interf_illustr}
\end{figure}

Let us consider a LoRa gateway which is perfectly synchronized to the signal of a desired user on which the signal of an interfering user is superimposed. Although the impact of inter-SF interference has been shown to be non-negligible~\cite{Croce2020}, it is quite different than the impact of same-SF interference, due to spreading gain. As a result, the two types of interference need to be analyzed separately. Due to the approximate orthogonality of different spreading factors, inter-SF interference results in a wide-band spectrum with low spectral density in the frequency domain after dechirping~\cite{Croce2020}. Therefore, inter-SF interference can be approximately treated as white noise and can be included in our model by properly adjusting the SNR. In this Section we only consider the case where the interfering packet has the same SF as the desired user, which is the most severe type of interference. To simplify the analysis, we consider only one interfering user. If multiple interfering users collide at the same time, the strongest user typically dominates the impact on the error rate.

Since LoRa uses the non-slotted ALOHA protocol for medium access control, the interfering signal ${y_{I}[n] = h_{I}x_{I}[n]}$ is neither synchronized to the desired user nor to the gateway. However, we assume that the gateway is perfectly synchronized to the desired user, using one of the known synchronization methods~\cite{Bernier2019,Xhonneux2019}, as shown in~\cite{Tapparel2020}. Moreover, we assume that any carrier frequency offset of the desired user has been perfectly estimated and compensated. However, the gateway can perform CFO estimation and compensation only for one user, therefore, there is no CFO estimation and compensation for the interfering user. Thus, we need to include the CFO of the interferer in order to obtain an accurate model.

As shown in~Fig.~\ref{fig:interf_illustr}, due to the lack of any time synchronization with the interferer, the interfering signal $x_{I}[n]$ will generally consist of parts of two distinct LoRa symbols, which we denote by $s_{I_{1}}$ and $s_{I_{2}}$. Following the notation of~\cite{Afisiadis2020}, let $\tau$ be the relative time offset between the first chip of the transmitted symbol of interest $s$ and the first chip of the interfering symbol $s_{I_{2}}$ (i.e., the first chip of the second interfering symbol $s_{I_{2}}$ starts $\tau$ chip durations \emph{after} the first chip of the desired symbol $s$). We note that the offset $\tau$ can be split into an integer part $L = \left \lfloor{\tau}\right \rfloor$, and a non-integer part $\lambda = \tau - \left \lfloor{\tau}\right \rfloor$. We consider that $\tau$ has a uniform distribution in $[0, N)$, due to the complete lack of synchronization. The discrete-time baseband equivalent equation\footnote{We note that the exact expression for $x_I[n]$, as given by~\eqref{eq:xI_fullModel}, \emph{requires} the use of~\eqref{eq:LoRa_symbol_twoEqs} for representing each one of the two interfering symbols $ s_{I_{1}} $ and $ s_{I_{2}}$. In our expression for $x_I[n]$ in~\cite[Eq. (21)]{Afisiadis2020}, we instead use~\eqref{eq:LoRa_symbol} for representing $ s_{I_{1}} $ and $ s_{I_{2}}$. Therefore, ~\cite[Eq. (21)]{Afisiadis2020} can only be considered as an approximation of the actual expression for $x_I[n]$ in the presence of non-integer time offsets $\tau$. In \cite{Afisiadis2020} we use the aforementioned simplification, but in the current work we present the exact expression for a more realistic analysis.} of $x_I[n]$ can be found using~\eqref{eq:LoRa_symbol_twoEqs} for $ s_{I_{1}} $ and $ s_{I_{2}}$, appropriately adjusted to include the time offset $\tau$
\begin{align} \label{eq:xI_fullModel}
x_{I}[n] & =
\begin{cases}
e^{j2\pi \left(\frac{(n + N-\tau)^{2}}{2N} + (n + N-\tau)\left(\frac{s_{I_{1}}}{N}-\frac{1}{2}\right)\right)}, & n \in \mathcal{N}_{L_1},\\
e^{j2\pi \left(\frac{(n + N-\tau)^{2}}{2N} + (n + N-\tau)\left(\frac{s_{I_{1}}}{N}-\frac{3}{2}\right)\right)}, & n \in \mathcal{N}_{L_2},\\
e^{j2\pi \left(\frac{(n - \tau)^{2}}{2N} + (n - \tau)\left(\frac{s_{I_{2}}}{N}-\frac{1}{2}\right)\right)}, & n \in \mathcal{N}_{L_3},\\
e^{j2\pi \left(\frac{(n - \tau)^{2}}{2N} + (n - \tau)\left(\frac{s_{I_{2}}}{N}-\frac{3}{2}\right)\right)}, & n \in \mathcal{N}_{L_4},
\end{cases}
\end{align}
where 
\begin{align}
\mathcal{N}_{L_1} &= \{0,\dots,\lceil \tau\rceil{-}s_{I_{1}}{-}1\}, \\
\mathcal{N}_{L_2} &= \{\max\left(\lceil \tau\rceil{-}s_{I_{1}},0\right),\dots,\lceil \tau\rceil{-}1\}, \\
\mathcal{N}_{L_3} &= \{\lceil \tau\rceil, \dots,\min\left(N{-}s_{I_{2}}{+}\lceil \tau\rceil{-}1,N{-}1\right)\}, \\
\mathcal{N}_{L_4} &= \{N{-}s_{I_{2}}{+}\lceil \tau\rceil, \dots,N{-}1\},
\end{align}
and where we define $\{a,\hdots,b\} = \emptyset$ if $b<a$.

Furthermore, let $ f_{c_I} $ be the carrier frequency used during up-conversion at the transmitter of the interfering user and $ f_{c} $ be the carrier frequency used during down-conversion at the gateway after alignment with the carrier frequency of the desired user. The carrier frequency offset of the interfering user is the difference \mbox{$\Delta f_{c} = f_{c_I} {-} f_{c}$}, while any frequency offset of the desired user is perfectly compensated. As a result, the corresponding signal model is
\begin{align}
y[n] & = hx[n] + h_{I}c_{I}[n]x_{I}[n] + z[n], \;\; n \in \mathcal{S}, \label{eq:lora_rx_int}
\end{align}
where $h$ is the channel gain between the user and the LoRa gateway, $x[n]$ is the transmitted signal, $h_{I}$ is the channel gain between the interferer and the gateway, $x_{I}[n]$ is the transmitted interfering signal, $ c_{I}[n] = e^{j2\pi (n{+}(m-1)N) \frac{\Delta f_{c}}{f_{s}}}$ is the CFO term affecting the $m$-th symbol in the interfering packet, and $z[n] \sim \mathcal{N}(0,\sigma^2)$ is AWGN. We note that, after despreading, the CFO translates into a time offset $\tau_{\text{cfo}} = \frac{\Delta f_{c}N}{f_{s}}$~\cite{Afisiadis2019c}, and similarly to $\tau$, the offset $\tau_{\text{cfo}}$ can be split into an integer part $L_{\text{cfo}} = \left \lfloor{\tau_{\text{cfo}}}\right \rfloor$, and a non-integer part $\lambda_{\text{cfo}} = \tau_{\text{cfo}} - \left \lfloor{\tau_{\text{cfo}}}\right \rfloor$. Since $|h| = 1$, the signal-to-interference ratio (SIR) can be defined as $\text{SIR} = \frac{1}{P_I}$, where $P_I=|h_I|^2$ is the received power of the interfering user at the gateway. The demodulation of $y[n]$ at the receiver yields
\begin{align}
\mathbf{Y}  
& = \text{DFT}\left(h\mathbf{x} \odot \mathbf{x}_{\text{ref}}^{*}\right) + \text{DFT}\left(h_I\mathbf{c}_{I} \odot \mathbf{x}_{I} \odot \mathbf{x}_{\text{ref}}^{*}\right) \nonumber\\ 
& + \text{DFT}\left(\mathbf{z} \odot \mathbf{x}_{\text{ref}}^{*}\right).
\end{align}
We call $\text{DFT}\left(h_I\mathbf{c}_{I} \odot \mathbf{x}_{I} \odot \mathbf{x}_{\text{ref}}^{*}\right) = \text{DFT}(\mathbf{y}_{I} \odot \mathbf{x}_{\text{ref}}^{*})$ the \emph{received interference pattern}. The received interference pattern depends on the time-domain interference signal $\mathbf{y}_{I}$, which is in turn a function of the interfering symbols $ s_{I_{1}} $, $ s_{I_{2}} $, the channel $h_I$, the interferer time offset $ \tau $, and the CFO between the interferer and the gateway.

\begin{figure}[tb] 
	\centering
	\subfloat[The magnitude of the received interference pattern]{%
		\includegraphics[width=0.5\textwidth]{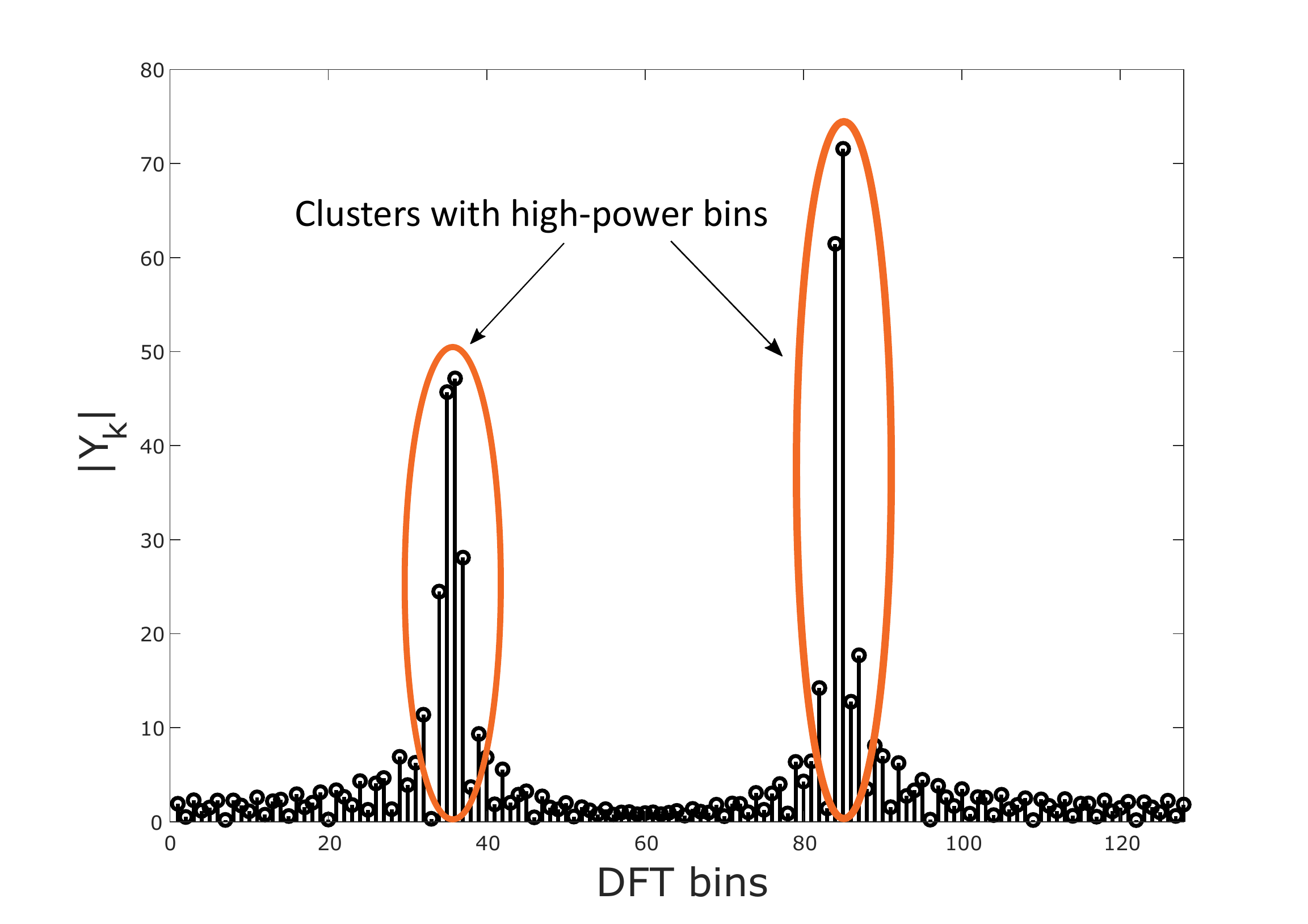}%
		\label{fig:RxInterfPattern}
	}\quad
	\subfloat[The complex plane representation of the received interference pattern]{%
		\includegraphics[width=0.5\textwidth]{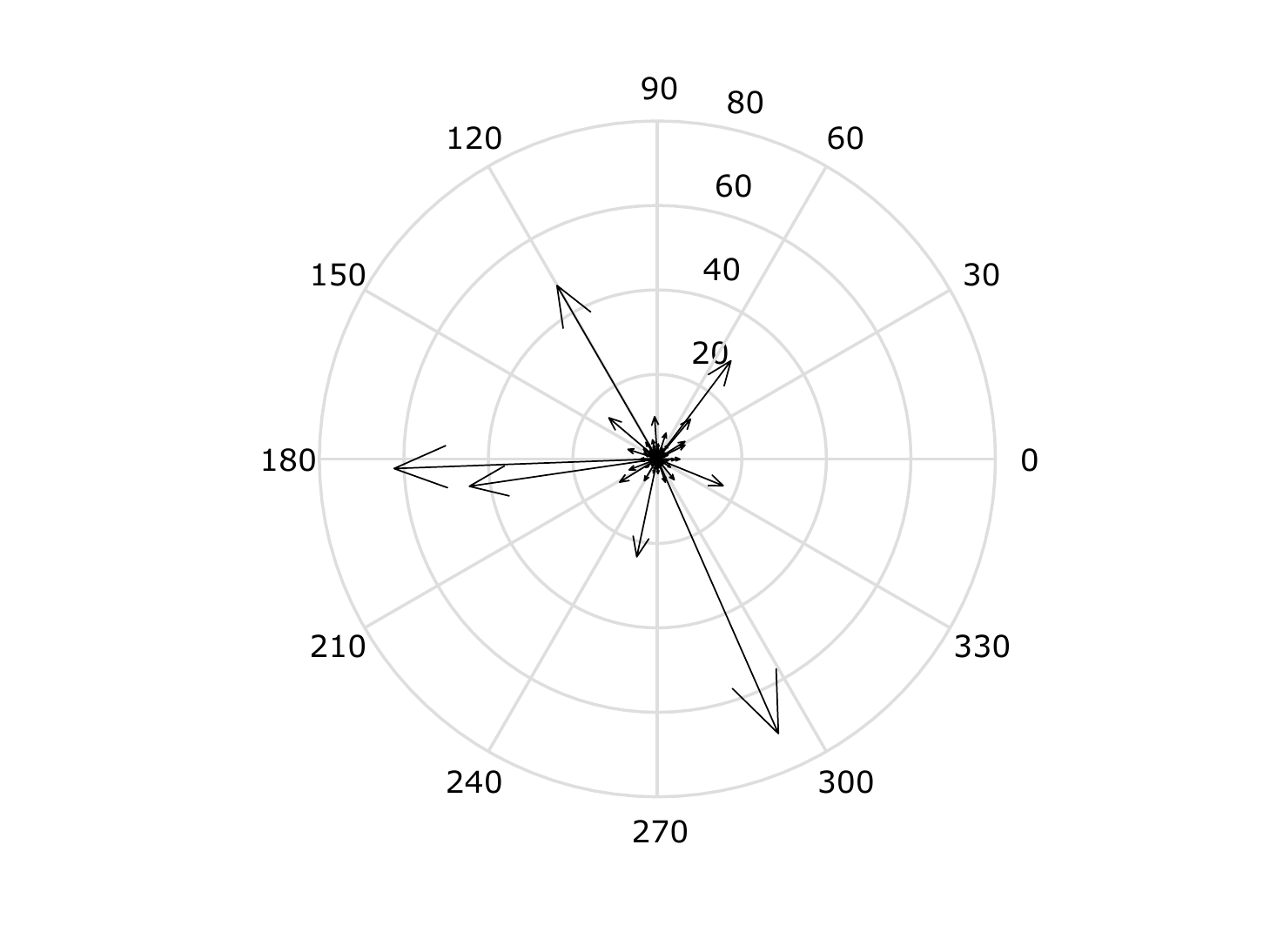}
		\label{fig:compass}
	}
	\caption{The received interference pattern for $\text{SF}=7$, $s_{I_{1}}=83$, $s_{I_{2}}=4$, $\tau=88.4$, $|h_I|=1$, and $\lambda_{\text{cfo}}=0.4$.}
	\label{fig:InterfPattern}
\end{figure}

We note that, in the presence of a fractional CFO, the received interference pattern is not just a scaled version of the pattern without CFO, but it has a different shape. A LoRa signal model in the presence of both time and frequency offsets has first been discussed in~\cite{Xhonneux2019} for the LoRa preamble. An example of a received interference pattern with CFO is shown in Fig.~\ref{fig:InterfPattern}. In Fig.~\ref{fig:RxInterfPattern}, we show the magnitude of the interference pattern across different bins after the DFT, while Fig.~\ref{fig:compass} shows also the phase in the complex plane. In coherent detection of LoRa, the received interference pattern has to be examined on the complex plane, since the symbol decision is performed after projection on the real axis instead of based only on the magnitude as in~\cite{Afisiadis2020}.

\begin{figure*}
\begin{align} \label{eq:ser_full}
P\left(\hat{s}{\neq} s|\tau_{\text{cfo}}\right) & = 1-\frac{1}{\sigma(2\pi)^{\frac{3}{2}}}\sum_{s{=}0}^{N{-}1}\sum _{s_{I_{1}}{=}0}^{N{-}1}\sum _{s_{I_{2}}{=}0}^{N{-}1} \int _{\tau {=} 0}^{N}\int_{\omega{=}0}^{2 \pi}\int_{y{=}0}^{{+}\infty} e^{{-}\frac{\left(y{-}\mu_{s}\right)^2}{2\sigma^{2}}} \prod_{\substack{k{=}1\\k\neq s}}^{N} F_{\mathcal{N}}(y;\mu_k,\sigma^{2}) dy d\omega d \tau
\end{align}
	\hrule
\end{figure*}

\subsection{Symbol Error Rate Under Same-SF Interference}

With the above described signal and interference model, we can now derive the SER expression following an approach that is similar to~\cite{Afisiadis2020}.

\subsubsection{Distribution of the Decision Metric}\label{sec:distdecmetr}
Let $ R_{k} $ be the value of the received interference pattern at frequency bin $k$.
For a specific combination of a symbol s and an interference pattern $\mathbf{y}_I$, the distribution of $ Y_k $ is
{
	\begin{align}
	Y_k {\sim}
	\begin{cases}
	\mathcal{CN}\left(|R_{k}|\cos\theta_{k} {+} j|R_{k}|\sin\theta_{k},2\sigma^{2} \right), & k \in \mathcal{S} / s \\
	\mathcal{CN}\left(N\cos\phi {+} |R_{k}|\cos\theta_{k} {+}\right. \nonumber \\
	{+} \left. j\left(N\sin\phi {+} |R_{k}|\sin\theta_{k}\right),2\sigma^{2} \right), & k = s, \\
	\end{cases}
	\end{align}
}%
where $\theta_{k} = \theta + \theta_{I_{k}}$ is the phase shift for bin k introduced by the interference channel and by the CFO. We note that $\theta$ is fixed for all symbols in a given packet, but changes for different packet transmissions, and is generally uniformly distributed in $[0,2\pi)$. However, the phase offset induced by the CFO of the interferer, $\theta_{I_{k}}$, is deterministic, but different for each bin $k$, and changes for each symbol in a packet. Since the received signal $\mathbf{U}$ is rotated by $-\phi$ due to the coherent detection, we define the phase shift between the interferer and the user as $\omega_{k} = \omega + \theta_{I_{k}}$, where $\omega = \theta - \phi$, which corresponds to relative phase shift of the interferer, after the rotation introduced by the coherent receiver. We note that since $\phi$ is fixed for each transmission, but generally uniformly distributed in $[0,2\pi)$, it holds that $\omega$, is also fixed for each transmission, but generally uniformly distributed in $[0,2\pi)$.

The demodulation metric $\Re(U_k) $ in the presence of interference, is therefore distributed as
\begin{align}
\Re(U_k) {\sim}
\begin{cases}
\mathcal{N}\left(|R_{k}|\cos\omega_{k} ,\sigma^{2} \right), & k \in \mathcal{S} / s \\
\mathcal{N}\left(N {+} |R_{k}|\cos\omega_{k},\sigma^{2} \right), & k = s.
\end{cases}
\end{align}

\subsubsection{Symbol Error Rate Expression}
For a given realization of a transmission, conditioning on both the phase $\phi$ of the channel of the desired user and the phase $\theta$ of the channel of the interfering user, can be replaced by conditioning on $\omega = \theta - \phi$. For notation simplicity, we denote the SER for a given symbol $s$, conditioned on $s_{I_{1}}$, $s_{I_{2}}$, the relative offset $\tau$, the equivalent offset due to the CFO $\tau_{\text{cfo}}$, and the phase difference $\omega$, as $ P\left(\hat{s}\neq s|s,s_{I_{1}},s_{I_{2}},\tau,\tau_{\text{cfo}},\omega\right) = P\left(\hat{s}{\neq} s|s,\mathbf{y}_{I},\omega\right)$. This error probability can be written as
\begin{align} \label{eq:formula_SER_noise_interf}
P\left(\hat{s}{\neq} s|s,\mathbf{y}_{I},\omega\right) & = 1{-} \frac{1}{\sigma\sqrt{2\pi}} \int_{y=0}^{+\infty} e^{{-}\frac{\left(y{-}\mu_{s}\right)^2}{2\sigma^{2}}} F_{\Re(U_{\max})} (y) dy,
\end{align}
where $\mu_{s} = N {+} |R_{s}|\cos\omega_{s}$, and $F_{\Re(U_{\max})} (y) = \prod_{\substack{k=1\\k\neq s}}^{N} F_{\mathcal{N}}(y;\mu_k,\sigma^{2})$, where $F_{\mathcal{N}}(y;\mu_k,\sigma^{2})$ denotes the CDF of a Gaussian distribution, and where $\mu_{k} = |R_{k}|\cos\omega_{k}$.
Following a similar reasoning as in~\cite{Afisiadis2020} for the non-coherent receiver, the full expression for $P\left(\hat{s}{\neq} s|\tau_{\text{cfo}}\right) $ in the case of a coherent receiver, conditioned only on the CFO value, is given in~\eqref{eq:ser_full}.

\section{Symbol Error Rate Bound and Approximation} \label{sec:bound_approx}

Unfortunately, the computational complexity for evaluating the expression in~\eqref{eq:ser_full} is prohibitive and numerical problems may also arise. Therefore, in this section, we derive a lower bound as well as an approximation based on this bound for efficiently evaluating~\eqref{eq:ser_full} in the practically relevant SNR operating regime of LoRa.

\subsection{Interference Patterns}
Let $V_{k}$ be the value of the received interference pattern at frequency bin $k$ after the rotation due to the coherent receiver, i.e., $V_{k} = e^{-j\phi}R_{k}$, assuming perfect knowledge of $\phi$. We first derive an explicit form for the real-axis projection of $V_{k}$, i.e., $\Re(V_k),~k\in\mathcal{S}$. Using the definition of the DFT and after some simple algebraic transformations, we obtain
\begin{align}
\Re(V_{k}) & = |h_{I}|\sum_{j=1}^{4}A_{k,j} \cos{\theta_{k,j}}, \label{eq:ReRk}
\end{align}
where $A_{k,j}$ and $\theta_{k,j}$ are given in the Appendix.

\subsection{Symbol Error Rate Bound}

\begin{figure*}
	\begin{align}
	\mathcal{D}_{1} &= \left\{\left[ N -\lfloor \tau{-}\tau_{\text{cfo}} \rceil + s_{I_{1}}-\frac{K-1}{2} \right]_N, \dots, \left[ N -\lfloor \tau{-}\tau_{\text{cfo}} \rceil + s_{I_{1}}+\frac{K-1}{2} \right]_N\right\}  \label{eq:D1}\\ 
	\mathcal{D}_{2} &= \left\{ \left[N -\lfloor \tau{-}\tau_{\text{cfo}} \rceil + s_{I_{2}}-\frac{K-1}{2} \right]_N, \dots, \left[N -\lfloor \tau{-}\tau_{\text{cfo}} \rceil + s_{I_{2}}+\frac{K-1}{2} \right]_N\right\} \label{eq:D2}
	\end{align}
	\hrule
\end{figure*}

We now wish to derive a bound for $P\left(\hat{s}{\neq} s|\tau_{\text{cfo}}\right)$, with the use of Q-functions, that can be evaluated more efficiently than~\eqref{eq:ser_full}. First, we assume that the maximum of $\Re(V_k)$, as depicted in Fig.~\ref{fig:bins_coh}, dominates the interference-induced SER. In particular, taking into consideration only the interfering bin with the maximum projection results in a lower bound for the probability of error. We need to search over all $N-1$ possible erroneous bins in order to evaluate the maximum projection
\begin{align}
  V_{k_{\max}} & = \max_{k\in \mathcal{S}/s }\left(\Re(V_{k})\right).\label{eq:maxProjection}
\end{align}

The decision metric for the bin of the maximum projection is Gaussian distributed with mean $V_{k_{\max}}$ and variance $\sigma^{2}$, while the decision metric for the desired bin $s$ is Gaussian distributed with mean $N + \Re(V_{s})$ and variance $\sigma^{2}$. Therefore, their difference also follows a Gaussian distribution with mean $V_{k_{\max}}-N-\Re(V_{s})$ and variance $2\sigma^{2}$, and a symbol error happens if this difference is positive. Thus, for a given transmitted symbol $s$, a particular realization of the user channel, and a particular realization of the interference pattern, i.e., for given $s$, $s_{I_{1}}$, $s_{I_{2}}$, $\tau$, $\tau_{\text{cfo}}$, and $\omega$, the probability that the interfering bin has higher energy than bin $s$ can be bounded from below by
\begin{align}
	P(\hat{s}\neq s|s,\mathbf{y}_{I},\omega) & \geq Q\left(\frac{N + \Re(V_{s}) -V_{k_{\max}}}{\sqrt{2\sigma^2}}\right),
\end{align}
where $\Re(V_{s})$ is the projection of the interference which lies on top of our desired symbol on bin $s$.
Since symbol s can take any of the $N$ possible values, the error rate after removing the conditioning on $s$ can be written as
\begin{align}
P(\hat{s}{\neq} s|\mathbf{y}_{I},\omega) & \geq \frac{1}{N} \sum _{s=0}^{N{-}1}Q\left(\frac{N + \Re(V_{s}) -V_{k_{\max}}}{\sqrt{2\sigma^2}}\right).\label{eq:Qmean}
\end{align}

Using~\eqref{eq:Qmean}, we can finally bound from below the interference-induced SER as
\begin{align}
P\left(\hat{s}{\neq} s|\tau_{\text{cfo}}\right)  & = \frac{2\pi}{N^3}\sum _{s_{I_1}=0}^{N{-}1}\sum _{s_{I_2}=0}^{N{-}1} \int _{0}^{2\pi} \int _{0}^{N} P(\hat{s}{\neq} s|\mathbf{y}_{I},\omega) d \tau d \omega. \label{eq:approxint}
\end{align}

We can further reduce the complexity for evaluating the interference-induced SER in~\eqref{eq:approxint} by reducing the complexity of evaluating $P(\hat{s}{\neq} s|\mathbf{y}_{I},\omega)$ using a less tight lower bound. Looking at~\eqref{eq:A1}--\eqref{eq:A4} in the Appendix, we observe that the DFT bins adjacent to bins $N -\lfloor \tau{-}\tau_{\text{cfo}} \rceil + s_{I_{1}}$ and $N -\lfloor \tau{-}\tau_{\text{cfo}} \rceil + s_{I_{2}}$ have considerably higher energy, compared to the rest of the bins, due to the combination of the fractional part of the misalignment and the fractional part of the CFO. An example of this can be observed in Fig.~\ref{fig:RxInterfPattern}.

Let $K$ define the number of most relevant bins in each cluster of high-power DFT bins, as depicted in Fig.~\ref{fig:RxInterfPattern}. Let $\mathcal{D} = \mathcal{D}_{1} \cup \mathcal{D}_{2}$, with $\mathcal{D}_{1}$ and $\mathcal{D}_{2}$ given by~\eqref{eq:D1}--\eqref{eq:D2},
where $|\mathcal{D}|=2K$. $\mathcal{D}_{1}$ and $\mathcal{D}_{2}$ are the sets of $K$ most relevant bins in the cluster around $N {-}\lfloor \tau{-}\tau_{\text{cfo}} \rceil {+} s_{I_{1}}$ and of $K$ most relevant bins in the cluster around $N {-}\lfloor \tau{-}\tau_{\text{cfo}} \rceil {+} s_{I_{2}}$, respectively, where we denote $[x]_y = x \mod y$. We note here that the two sets $\mathcal{D}_{1}$ and $\mathcal{D}_{2}$ of the most relevant bins can have overlapping bins. We are interested in treating the set of most relevant bins $\mathcal{D} = \mathcal{D}_{1} \cup \mathcal{D}_{2}$ separately from the rest of the bins. As can be seen in Fig.~\ref{fig:compass} and Fig.~\ref{fig:bins_coh}, there is a high probability that the maximum projection $V_{k_{\max}}$ will occur due to one of these most relevant bins that belong to $\mathcal{D}$. Increasing $K$ (i.e., increasing also the cardinality of the set $\mathcal{D}$) increases the number of cases in which the bin with the maximum projection will be included in the set $\mathcal{D}$. We reduce the set of bins we consider from $\mathcal{S}$ with cardinality $|\mathcal{S}|=N$, to $\mathcal{D}$ with much smaller cardinality $|\mathcal{D}|<<N$ which does not explicitly scale with SF. This way the search in equation~\eqref{eq:maxProjection} is only conducted over $\mathcal{D}$, instead of $\mathcal{S}$ as
\begin{align}
V_{k_{\max}} & = \max_{k\in \mathcal{D}/s }\left(\Re(V_{k})\right).\label{eq:maxProjection_D}
\end{align}
By using~\eqref{eq:maxProjection_D} instead of~\eqref{eq:maxProjection} in~\eqref{eq:Qmean}, we obtain a lower bound which can be evaluated with lower complexity.

\subsection{Symbol Error Rate Approximation}

We now wish to derive an approximation for $P\left(\hat{s}{\neq} s|\tau_{\text{cfo}}\right)$ that can be evaluated with even less computational complexity than~\eqref{eq:Qmean}--\eqref{eq:approxint}.

The impact of the interference at bin $s$ is significant only when $s\in \mathcal{D}$. For all cases where $s\notin \mathcal{D}$ the impact of the interference at bin $s$ is negligible even if it aligns with the real axis. By distinguishing the $N$ bins in two sets, namely set $\mathcal{D}$ where we account for the interference impact at bin $s$, and set $\mathcal{S}/\mathcal{D}$  where we do \emph{not} account for the interference impact at bin $s$,~\eqref{eq:Qmean} can be approximated as
\begin{align}
P(\hat{s}{\neq} s|\mathbf{y}_{I},\omega) & \approx \frac{1}{N} \left(\sum _{s\in \mathcal{D}}Q\left(\frac{N + \Re(V_{s}) -V_{k_{\max}}}{\sqrt{2\sigma^2}}\right) \right. \nonumber \\
& + \left. (N-|\mathcal{D}|)Q\left(\frac{N -V_{k_{\max}}}{\sqrt{2\sigma^2}}\right) \right). \label{eq:Qmean_setD}
\end{align}
Using~\eqref{eq:Qmean_setD}, $P(\hat{s}{\neq} s|\mathbf{y}_{I},\omega)$ only requires the evaluation of $|\mathcal{D}| + 1$ Q-functions, instead of $N$ Q-functions needed for~\eqref{eq:Qmean}, which again is specifically advantageous for large SFs (corresponding to large $N$). We show in the results, that choosing a very small $K$, e.g., $K=5$, i.e., $|\mathcal{D}| = 10$, is sufficient for accurate results, for all examined SFs. Finally, the approximation in~\eqref{eq:Qmean_setD} can be replaced directly in~\eqref{eq:approxint}.

The integral over the offset $\tau$ in~\eqref{eq:approxint} can be evaluated numerically by discretizing the interval $[0,N)$ with a step size $\epsilon$. Moreover, the integral over $\omega$ can be evaluated by discretizing the interval $[0,2\pi)$ with a step size $\rho$. In the results, we show that a discretization step size choice of $\epsilon = \frac{1}{5}$ and $\rho = \frac{\pi}{2}$ provides a low-complexity, yet accurate, evaluation of~\eqref{eq:approxint}.

In the AWGN-limited regime (i.e., at low SNR), the above approximation eventually becomes inaccurate, since all bins have similar values and no single bin projection dominates the error rate. To fix this issue, let $P^{(N)}(\hat{s}{\neq} s)$ denote the SER under AWGN given in~\eqref{eq:formula_SER_coh_only_noise_finalform}, which can be efficiently evaluated using the approximation in~\eqref{eq:SER_AWGN_approx}, and let $P^{(I)}(\hat{s}{\neq} s|\tau_{\text{cfo}})$ be the interference-driven SER from~\eqref{eq:approxint}. Then, a final estimate of the average SER that is accurate also in the low-SNR regime can be approximated by
\begin{align}
P\left(\hat{s}{\neq} s|\tau_{\text{cfo}}\right)  & \approx P^{(N)}(\hat{s}{\neq} s) {+} \left(1{-}P^{(N)}(\hat{s}{\neq} s)\right)P^{(I)}\left(\hat{s}{\neq} s|\tau_{\text{cfo}}\right). \label{eq:approxfinal}
\end{align}

\section{Coherent LoRa Frame Error Rate \\Under Same-SF Interference} \label{sec:FER_interf}

\begin{figure*}
	\begin{align}
	P(\hat{\mathbf{s}} \neq  \mathbf{s}|F_{I},\tau_{\text{cfo}})  & = \frac{1}{N}\int _{0}^{N}\left(1-\left(1- P(\hat{s} \neq s|\tau,\tau_{\text{cfo}})\right)^{F_{I}}\left(1-P^{(N)}(\hat{s}\neq s)\right)^{F-F_{I}}\right) d\tau
	\label{eq:FER_coh_partialCollision}
	\end{align}
	\hrule
\end{figure*}

In system-level simulator environments the FER is generally of greater practical interest than the SER, because system-level simulators, typically take decisions on a frame-by-frame basis. Therefore, in this section we derive expressions to calculate the FER of a coherent uncoded LoRa system. These expression can be used for the LoRa modes that use channel codes of rates $\nicefrac{4}{5}$, and $\nicefrac{4}{6}$, which have error-detection, but no error-correction capabilities, as well as for the uncoded mode. The extension of the analysis to a coded LoRa system can be done following the methodology of~\cite{Afisiadis2019c}.

\subsubsection{FER for the overlapping portion of the packets}

We assume perfect frame synchronization for the desired user, even under the impact of interference. Generally, due to the time offset between the desired and the interfering frame, only part of the desired frame is affected by interference. In the following, we consider only the part of the frame that is affected by interference. This way, our derived expression is a straightforward basis for computing FERs for any partial collision of two frames by splitting the frame under consideration into a collision interval which is affected by interference and a second part that is affected only by AWGN. The averaging over all possible relative positions of the packets through Monte Carlo simulations is then inherently done by the network simulator itself.

A collision interval of $F$ LoRa symbols is denoted by the vector $\mathbf{s}$ and the corresponding estimated symbols at the receiver are denoted by the vector $\mathbf{\hat{s}}$. The FER is therefore $P(\hat{\mathbf{s}}\neq \mathbf{s})$. We note that the expression for the SER derived in Section~\ref{sec:SER_interf} can not be used as is for the evaluation of the FER  because it includes an expectation over the random variables $\tau$ and $\omega$, while all symbols in a frame experience the same $\tau$ and the same $\omega$ over the course of the frame. However, the CFO of the interferer results in a continuous rotation of the samples in the interfering packet. Therefore, for sufficiently long frames, we consider the phase of each symbol (as determined by the CFO) as random and allow the expectation over $\omega$ to be included on a symbol level. The frame error rate $P(\hat{\mathbf{s}}\neq \mathbf{s})$ can thus be expressed as
	\begin{align}
	P(\hat{\mathbf{s}} \neq  \mathbf{s}|\tau_{\text{cfo}})  & = \frac{1}{N}\int _{0}^{N}\left(1-(1-P(\hat{s} \neq s|\tau,\tau_{\text{cfo}}))^F\right) d\tau. \label{eq:FER_coh}
	\end{align}
	where $P(\hat{s}\neq s|\tau,\tau_{\text{cfo}})$ is
	\begin{align}
	P(\hat{s}\neq s|\tau,\tau_{\text{cfo}}) & = \frac{2\pi}{N^2}\sum _{s_{I_1}=0}^{N{-}1}\sum _{s_{I_2}=0}^{N{-}1}\int _{0}^{2\pi} P\left(\hat{s}{\neq} s|\mathbf{y}_{I},\omega\right) d \omega. \label{eq:FER_coh_givenTau}
	\end{align}
The frame error rate in~\eqref{eq:FER_coh} can be considered as an exact expression if we evaluate~\eqref{eq:FER_coh_givenTau} using~\eqref{eq:formula_SER_noise_interf} (after taking the expectation over $s$), or as a lower bound if we evaluate~\eqref{eq:FER_coh_givenTau} using~\eqref{eq:Qmean}, or as an approximation if we evaluate~\eqref{eq:FER_coh_givenTau} using~\eqref{eq:Qmean_setD}.

\subsubsection{Average FER over different overlapping portions of the packets}

In the following, we also present an expression for the average FER over all possible relative positions of the colliding packets. This expression is useful in cases where the averaging over all possible relative positions of the packets is \emph{not} inherently done by a network simulator. To this end, we consider that the interfering frame has the same length as the desired frame. This assumption is taken only for clarity of the presentation and does not prevent the generalization of the results to any interfering frame length. The difference to~\eqref{eq:FER_coh} is that, due to the time offset between the frames, only part of the frame of interest is affected by interference, and an average of all collision intervals is considered.

As in \cite{Afisiadis2020}, let $F_{I} \in \{1, \dots, F \} $, denote the number of symbols in the frame that are affected by the interfering frame. The value of $F_{I}$ depends on the relative position of the two frames. For the average FER, we consider the expectation over all the possible relative positions of the two frames. We note that, except in the case of perfect alignment between the frame of interest and the interference, there always exists one symbol that is only partially affected by interference. Similarly to \cite{Afisiadis2020}, we consider the partially-affected symbol as fully-affected by interference, thus including it in $F_{I}$. The frame error rate is now given by~\eqref{eq:FER_coh_partialCollision}
where $P(\hat{s}\neq s|\tau,\tau_{\text{cfo}})$ is given by~\eqref{eq:FER_coh_givenTau}, and $P^{(N)}(\hat{s}\neq s)$ is the SER under AWGN given in~\eqref{eq:formula_SER_coh_only_noise_finalform} (which can be evaluated efficiently using the approximation in~\eqref{eq:SER_AWGN_approx}).

Finally, we take the expectation over all possible values of $ F_{I} $ and we obtain the final expression for the average FER over different collision intervals
\begin{align}
P_{\text{av}}(\hat{\mathbf{s}} \neq  \mathbf{s}|\tau_{\text{cfo}})   & \approx \frac{1}{F}\sum _{F_{I}=1}^{F} P(\hat{\mathbf{s}} \neq  \mathbf{s}|F_{I},\tau_{\text{cfo}}).\label{eq:approx_averageFERcoh}
\end{align}

\section{Results} \label{sec:results}

In this section, we provide numerical results for the SER and the FER of a coherent LoRa receiver with same-SF interference. We compare the performance against the non-coherent receiver under the same interference conditions. Moreover, we compare the derived approximations against Monte Carlo simulations to show their accuracy. We note that for the Monte Carlo simulations, the interferer time offset $ \tau $ is simulated by oversampling~\eqref{eq:LoRa_symbol_twoEqs} to create an accurate interference signal, concatenating the oversampled symbols, applying the appropriate offset, and downsampling the received signal to obtain $N=2^{\text{SF}}$ samples for $\mathbf{x}_{I}$.

\begin{figure*} 
	\centering 
	\begin{minipage}[t]{0.48\textwidth} 
		\begin{tikzpicture}

	\small

	\begin{semilogyaxis}[
		width = \figurewidth\columnwidth,
		height = \figureheight\columnwidth,
		xlabel = {SNR (dB)},
		ylabel = {Frame Error Rate},
		label style={font=\small},
    tick label style={font=\footnotesize},
		ylabel near ticks,
		xlabel near ticks,
		xmin = -15, xmax = 15,
		ymin = 1e-2, ymax = 1,
		grid = both,
		legend image post style={scale=0.6},
	]


		\addplot[Set1-7-1, thick, solid, mark=star, mark options={scale=1.2}] table[x index=0, y index = 1] {figs/data/PER_MC_SF7_F20_Fpr8_Pi0.00_lambda0.00_Coh0_Estim0_OS10_phOffset1_2Eqs_1_0.dat};
		\label{cfo0_MCnoncoh}
		\addplot[Set1-7-1, thick, solid, mark=diamond*, mark options={scale=1.3}] table[x index=0, y index = 1] {figs/data/PER_MC_SF7_F20_Fpr8_Pi0.00_lambda0.50_Coh0_Estim0_OS10_phOffset1_2Eqs_1_0.dat};
		\label{cfo05_MCnoncoh}
		
		\addplot[black, thick, dotted, mark=*, mark options={scale=0.6, solid}] table[x index=0, y index = 1] {figs/data/PER_APP_FER_NONcoh_SF7_F20_Pi0.00_tauCFO0.00_OS5_newCorrected_window5_noOmega.dat};
		\label{cfo0_APPnoncoh}
		\addplot[black, thick, dotted, mark=*, mark options={scale=0.6, solid}] table[x index=0, y index = 1] {figs/data/PER_APP_FER_NONcoh_SF7_F20_Pi0.00_tauCFO0.50_OS5_newCorrected_window3_s2_comb8.dat};
		\label{cfo05_APPnoncoh}


		\addplot[Set1-7-2, thick, solid, mark=*, mark options={scale=1.2}] table[x index=0, y index = 1] {figs/data/PER_MC_SF7_F20_Fpr8_Pi0.00_lambda0.00_Coh1_Estim0_OS10_phOffset1_2Eqs_1_0.dat};
		\label{cfo0_MCcoh}
		\addplot[Set1-7-2, thick, solid, mark=triangle*, mark options={scale=1.3}] table[x index=0, y index = 1] {figs/data/PER_MC_SF7_F20_Fpr8_Pi0.00_lambda0.50_Coh1_Estim0_OS10_phOffset1_2Eqs_1_0.dat};
		\label{cfo05_MCcoh}
		 
		\addplot[black, thick, dotted, mark=*, mark options={scale=0.6, solid}] table[x index=0, y index = 1] {figs/data/PER_APP_FER_coh_SF7_F20_Pi0.00_tauCFO0.00_OS5_newCorrected_window5.dat};
		\label{cfo0_APPcoh}
		\addplot[black, thick, dotted, mark=*, mark options={scale=0.6, solid}] table[x index=0, y index = 1] {figs/data/PER_APP_FER_coh_SF7_F20_Pi0.00_tauCFO0.50_OS5_newCorrected_window5.dat};
		\label{cfo05_APPcoh}

		\node [draw,fill=white,inner sep=2pt] at (rel axis cs: 0.23,0.15) {
		\tiny
		\begin{tabular}{lcc}
			\multicolumn{3}{c}{\scriptsize\underline{Coherent SF$=7$}} \\
											& MC 	             &    Approx.\\
			$\lambda_{\text{cfo}} = 0$:	 	& \ref{cfo0_MCcoh}   &    \ref{cfo0_APPcoh}       \\
			$\lambda_{\text{cfo}} = 0.5$: 	& \ref{cfo05_MCcoh}  &    \ref{cfo05_APPcoh} 
		\end{tabular}};

		\node [draw,fill=white,inner sep=2pt] at (rel axis cs: 0.77,0.85) {
		\tiny
		\begin{tabular}{lcc}
			\multicolumn{3}{c}{\scriptsize\underline{Non-coherent SF$=7$} } \\
								& MC 	                           &    Approx. \\ 
			$\lambda_{\text{cfo}} = 0$:	 	& \ref{cfo0_MCnoncoh}  &    \ref{cfo0_APPnoncoh} 	 \\
			$\lambda_{\text{cfo}} = 0.5$: 	& \ref{cfo05_MCnoncoh} &    \ref{cfo05_APPnoncoh}
		\end{tabular}};

	\end{semilogyaxis}

\end{tikzpicture}%
		\caption{Frame error rate of the coherent and non-coherent receiver for two different values of CFO, and a packet of length $F=20$ LoRa symbols under AWGN and same-SF interference for $\text{SF}=7$ and $P_I = 0$ dB.
		}
		\label{fig:ferSIR0dB}
	\end{minipage}
	$\quad$
	\begin{minipage}[t]{0.48\textwidth} 
		\begin{tikzpicture}

	\small

	\begin{semilogyaxis}[
		width = \figurewidth\columnwidth,
		height = \figureheight\columnwidth,
		xlabel = {SNR (dB)},
		ylabel = {Frame Error Rate},
		label style={font=\small},
    tick label style={font=\footnotesize},
		ylabel near ticks,
		xlabel near ticks,
		xmin = -12, xmax = 0,
		ymin = 1e-4, ymax = 1,
		grid = both,
		legend image post style={scale=0.6},
	]


		\addplot[Set1-7-1, thick, solid, mark=star, mark options={scale=1.2}] table[x index=0, y index = 1] {figs/data/PER_MC_SF7_F20_Fpr8_Pi-3.00_lambda0.00_Coh0_Estim0_sigmaPhi0.00_OS10_phOffset1_CFOJitter0_JittPerc0.00_2Eqs1_0.dat};
		\label{SIR3dB_cfo0_MCnoncoh}
		\addplot[Set1-7-1, thick, solid, mark=diamond*, mark options={scale=1.3}] table[x index=0, y index = 1] {figs/data/PER_MC_SF7_F20_Fpr8_Pi-3.00_lambda0.50_Coh0_Estim0_OS10_phOffset1_CFOJitter0_JittPerc0.10_2Eqs1_0.dat};
		\label{SIR3dB_cfo05_MCnoncoh}
		
		\addplot[black, thick, dotted, mark=*, mark options={scale=0.6, solid}] table[x index=0, y index = 1] {figs/data/PER_APP_FER_NONcoh_SF7_F20_Pi-3.00_tauCFO0.00_OS5_newCorrected_window5_s2_comb8_noOmega.dat};
		\label{SIR3dB_cfo0_APPnoncoh}
		\addplot[black, thick, dotted, mark=*, mark options={scale=0.6, solid}] table[x index=0, y index = 1] {figs/data/PER_APP_FER_NONcoh_SF7_F20_Pi-3.00_tauCFO0.50_OS5_newCorrected_window5_s2_comb8.dat};
		\label{SIR3dB_cfo05_APPnoncoh}


		\addplot[Set1-7-2, thick, solid, mark=*, mark options={scale=1.2}] table[x index=0, y index = 1] {figs/data/PER_MC_SF7_F20_Fpr8_Pi-3.00_lambda0.00_Coh1_Estim0_sigmaPhi0.00_OS10_phOffset1_CFOJitter0_JittPerc0.00_2Eqs1_0.dat};
		\label{SIR3dB_cfo0_MCcoh}
		\addplot[Set1-7-2, thick, solid, mark=triangle*, mark options={scale=1.3}] table[x index=0, y index = 1] {figs/data/PER_MC_SF7_F20_Fpr8_Pi-3.00_lambda0.50_Coh1_Estim0_OS10_phOffset1_CFOJitter0_JittPerc0.10_2Eqs1_filt_0.dat};
		\label{SIR3dB_cfo05_MCcoh}
		 
		\addplot[black, thick, dotted, mark=*, mark options={scale=0.6, solid}] table[x index=0, y index = 1] {figs/data/PER_APP_FER_coh_SF7_F20_Pi-3.00_tauCFO0.00_OS5_newCorrected_window5_s2_comb8.dat};
		\label{SIR3dB_cfo0_APPcoh}
		\addplot[black, thick, dotted, mark=*, mark options={scale=0.6, solid}] table[x index=0, y index = 1] {figs/data/PER_APP_FER_coh_SF7_F20_Pi-3.00_tauCFO0.50_OS5_newCorrected_window5_s2_comb8.dat};
		\label{SIR3dB_cfo05_APPcoh}

		\node [draw,fill=white,inner sep=2pt] at (rel axis cs: 0.23,0.15) {
		\tiny
		\begin{tabular}{lcc}
			\multicolumn{3}{c}{\scriptsize\underline{Coherent SF$=7$}} \\
											& MC 	             &    Approx.\\
			$\lambda_{\text{cfo}} = 0$:	 	& \ref{SIR3dB_cfo0_MCcoh}   &    \ref{SIR3dB_cfo0_APPcoh}       \\
			$\lambda_{\text{cfo}} = 0.5$: 	& \ref{cfo05_MCcoh}  &    \ref{SIR3dB_cfo05_APPcoh} 
		\end{tabular}};

		\node [draw,fill=white,inner sep=2pt] at (rel axis cs: 0.77,0.85) {
		\tiny
		\begin{tabular}{lcc}
			\multicolumn{3}{c}{\scriptsize\underline{Non-coherent SF$=7$} } \\
								& MC 	                           &    Approx. \\ 
			$\lambda_{\text{cfo}} = 0$:	 	& \ref{SIR3dB_cfo0_MCnoncoh}  &    \ref{SIR3dB_cfo0_APPnoncoh} 	 \\
			$\lambda_{\text{cfo}} = 0.5$: 	& \ref{SIR3dB_cfo05_MCnoncoh} &    \ref{SIR3dB_cfo05_APPnoncoh}
		\end{tabular}};

	\end{semilogyaxis}

\end{tikzpicture}%
		\caption{Frame error rate of the coherent and non-coherent receiver for two different values of CFO, and a packet of length $F=20$ LoRa symbols under AWGN and same-SF interference for $\text{SF}=7$ and $P_I = -3$ dB.
		}
		\label{fig:ferSIR3dB}
	\end{minipage}
\end{figure*}

\subsection{Error Rates}

In Fig.~\ref{fig:ferSIR0dB}, we show the results for the FER of a LoRa receiver for $\text{SF} = 7$, for a packet of length $F=20$ LoRa symbols, under the effect of same-SF interference with an SIR of $0$ dB, for $\lambda_{\text{cfo}} \in \{0, 0.5\}$, and for both a coherent and a non-coherent receiver. Results of both our model of~\eqref{eq:FER_coh} (black dotted lines), where we evaluate~\eqref{eq:FER_coh_givenTau} using the approximation in~\eqref{eq:Qmean_setD}, and of the Monte Carlo simulation (colored lines) are shown. We can see that the curves derived in our analysis match the Monte Carlo simulation curves very well, showing the accuracy of the corresponding approximations. We observe that both for the coherent and the standard non-coherent receiver, the fractional CFO of the interferer has a significant impact on the error rate. It is interesting to see that for lower values of SNR the fractional CFO of the interferer is beneficial for the error rate, but at high SNRs an interferer with a higher fractional CFO value results in a higher error floor. We note that the same behavior appears also for the non-coherent receiver, but at higher SNRs.

In Fig.~\ref{fig:ferSIR3dB}, we show the same results as in Fig.~\ref{fig:ferSIR0dB}, but for an SIR of $3$~dB. We note that the general trend of the curves is similar to the trend in Fig.~\ref{fig:ferSIR0dB}. However, we note that the results for $\text{SIR}=3$~dB show a slightly smaller (but still significant) impact of the CFO of the interferer on the error rate compared to the $\text{SIR}=0$~dB case depicted in Fig.~\ref{fig:ferSIR0dB}. We conclude that, since the CFO of the interferer is always present, it needs to be included in the model in order to examine its impact, which is not negligible. The inclusion of the CFO of the interferer in the model, and the discussion on its impact on the error rate of both coherent and non-coherent receivers, is more realistic than the model in~\cite{Afisiadis2020} for non-coherent receivers, where the CFO of the interferer was not included.

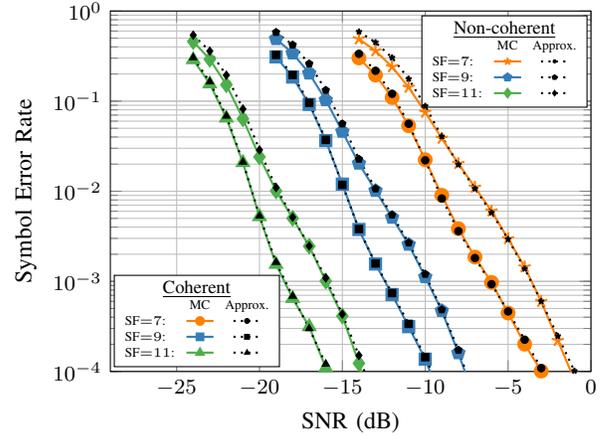
\begin{figure}[t]
	\centering
	\begin{tikzpicture}

	\small

	\begin{semilogyaxis}[
		width = \figurewidth\columnwidth,
		height = \figureheight\columnwidth,
		xlabel = {SNR (dB)},
		ylabel = {Symbol Error Rate},
		label style={font=\small},
    tick label style={font=\footnotesize},
		ylabel near ticks,
		xlabel near ticks,
		xmin = -29, xmax = 0,
		ymin = 1e-4, ymax = 1,
		grid = both,
		legend image post style={scale=0.6},
	]


		\addplot[Set1-7-5, thick, solid, mark=star, mark options={scale=1.2}] table[x index=0, y index = 1] {figs/data/PER_RES_SF7_Pi-3.00_OS10_phaseOffset1_coh0_twoEqs1_IoT_0.dat};
		\label{SF7MCnonCoh}
		\addplot[Set1-7-2, thick, solid, mark=pentagon*, mark options={scale=1.2}] table[x index=0, y index = 1] {figs/data/PER_RES_SF9_Pi-3.00_OS10_phaseOffset1_coh0_twoEqs1_IoT_0.dat};
		\label{SF9MCnonCoh}
		\addplot[Set1-7-3, thick, solid, mark=diamond*, mark options={scale=1.3}] table[x index=0, y index = 1] {figs/data/PER_RES_SF11_Pi-3.00_OS10_phaseOffset1_coh0_twoEqs1_IoT_0.dat};
		\label{SF11MCnonCoh}

		\addplot[black, thick, dotted, mark=star, mark options={scale=0.6, solid}] table[x index=0, y index = 1] {figs/data/APP_NONcoh_SF7_Pi-3.00_OS5.dat};
		\label{SF7APPnonCoh}
		\addplot[black, thick, dotted, mark=pentagon*, mark options={scale=0.6, solid}] table[x index=0, y index = 1] {figs/data/APP_SF9_Pi-3.00_OS3.dat};
		\label{SF9APPnonCoh}
		\addplot[black, thick, dotted, mark=diamond*, mark options={scale=0.6, solid}] table[x index=0, y index = 1] {figs/data/APP_SF11_Pi-3.00_OS3.dat};
		\label{SF11APPnonCoh}


		\addplot[Set1-7-5, thick, solid, mark=*, mark options={scale=1.2}] table[x index=0, y index = 1] {figs/data/PER_RES_SF7_Pi-3.00_OS10_phaseOffset1_coh1_twoEqs1_IoT_0.dat};
		\label{SF7MCCoh}
		\addplot[Set1-7-2, thick, solid, mark=square*, mark options={scale=1.05}] table[x index=0, y index = 1] {figs/data/PER_RES_SF9_Pi-3.00_OS10_phaseOffset1_coh1_twoEqs1_IoT_0.dat};
		\label{SF9MCCoh}
		\addplot[Set1-7-3, thick, solid, mark=triangle*, mark options={scale=1.3}] table[x index=0, y index = 1] {figs/data/PER_RES_SF11_Pi-3.00_OS10_phaseOffset1_coh1_twoEqs1_IoT_0.dat};
		\label{SF11MCCoh}

		\addplot[black, thick, dotted, mark=*, mark options={scale=0.6, solid}] table[x index=0, y index = 1] {figs/data/APP_coh_SF7_Pi-3.00_OS5.dat};
		\label{SF7APPCoh}
		\addplot[black, thick, dotted, mark=square*, mark options={scale=0.6, solid}] table[x index=0, y index = 1] {figs/data/APP_coh_SF9_Pi-3.00_OS5.dat};
		\label{SF9APPCoh}
		\addplot[black, thick, dotted, mark=triangle*, mark options={scale=0.6, solid}] table[x index=0, y index = 1] {figs/data/APP_coh_SF11_Pi-3.00_OS5.dat};
		\label{SF11APPCoh}

		\node [draw,fill=white,inner sep=2pt] at (rel axis cs: 0.18,0.14) {
		\tiny
		\begin{tabular}{lcc}
			\multicolumn{3}{c}{\scriptsize\underline{Coherent}} \\
								& MC 								& Approx. \\
			SF${=}7$:	 	& \ref{SF7MCCoh} 	&	\ref{SF7APPCoh} \\
			SF${=}9$: 	& \ref{SF9MCCoh} & \ref{SF9APPCoh} \\
			SF${=}11$: 	& \ref{SF11MCCoh} & \ref{SF11APPCoh}
		\end{tabular}};

		\node [draw,fill=white,inner sep=2pt] at (rel axis cs: 0.82,0.86) {
		\tiny
		\begin{tabular}{lcc}
			\multicolumn{3}{c}{\scriptsize\underline{Non-coherent} } \\
								& MC 								& Approx. \\
			SF${=}7$:	 	& \ref{SF7MCnonCoh} 	&	\ref{SF7APPnonCoh} \\
			SF${=}9$: 	& \ref{SF9MCnonCoh} & \ref{SF9APPnonCoh} \\
			SF${=}11$: 	& \ref{SF11MCnonCoh} & \ref{SF11APPnonCoh}
		\end{tabular}};

	\end{semilogyaxis}

\end{tikzpicture}%
	\caption{Symbol error rate of the coherent and non-coherent LoRa receiver under AWGN and same-SF interference for $\text{SF}
		\in \left\{7,9,11\right\}$ and $P_I = {-}3$ dB. The approximations for the coherent and the non-coherent case are shown with black dotted lines.}
	\label{fig:serint}
\end{figure}

In Fig.~\ref{fig:serint}, we show the results of a Monte Carlo simulation for the SER of a LoRa user for $\text{SF} \in \left\{7,9,11\right\}$ and for an SIR of 3~dB, using the coherent receiver described in this work, as well as the non-coherent receiver described in~\cite{Afisiadis2020}. The corresponding analytical approximations for the coherent and non-coherent receivers are shown as well. We observe that the coherent receiver has a significant performance gain of up to $2.5$~dB for some SNR values compared to the non-coherent receiver. In~\cite{Elshabrawy2019b,Marquet2019,Baruffa2019}, for the AWGN-only case, a difference of around $0.7$~dB was shown between the coherent and the non-coherent receiver. It is, therefore, very interesting to see that using a coherent receiver seems to be much more beneficial under same-SF interference  than in the AWGN-only case. Finally, we observe that the low-complexity SER expression in~\eqref{eq:approxint}, using the approximation in~\eqref{eq:Qmean_setD}, with $|\mathcal{D}| = 10$ and discretization steps $\rho = \frac{\pi}{2}$ and $\epsilon = \frac{1}{5}$ is already very accurate.

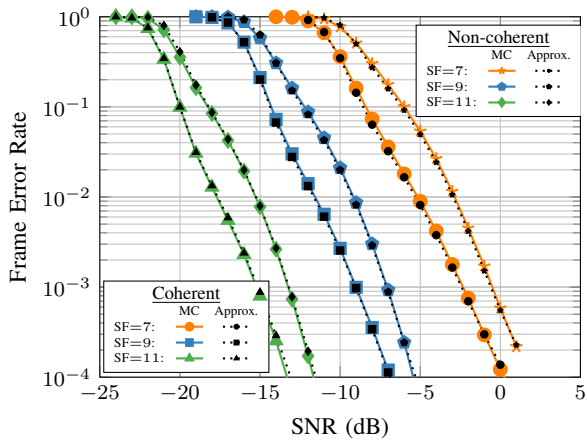
\begin{figure}[t]
	\centering
	\begin{tikzpicture}

	\small

	\begin{semilogyaxis}[
		width = \figurewidth\columnwidth,
		height = \figureheight\columnwidth,
		xlabel = {SNR (dB)},
		ylabel = {Frame Error Rate},
		label style={font=\small},
    tick label style={font=\footnotesize},
		ylabel near ticks,
		xlabel near ticks,
		xmin = -25, xmax = 5,
		ymin = 1e-4, ymax = 1,
		grid = both,
		legend image post style={scale=0.6},
	]


		\addplot[Set1-7-5, thick, solid, mark=star, mark options={scale=1.2}] table[x index=0, y index = 1] {figs/data/PER_MC_SF7_F20_Fpr8_Pi-3.00_lambda0.00_Coh0_Estim0_sigmaPhi0.00_OS10_phOffset1_CFOJitter0_JittPerc0.00_2Eqs1_0.dat};
		\label{SF7ferMCnoncoh}
		\addplot[Set1-7-2, thick, solid, mark=pentagon*, mark options={scale=1.2}] table[x index=0, y index = 1] {figs/data/PER_MC_SF9_F20_Fpr8_Pi-3.00_lambda0.00_Coh0_Estim0_sigmaPhi0.00_OS10_phOffset1_CFOJitter0_JittPerc0.00_2Eqs1_0.dat};
		\label{SF9ferMCnoncoh}
		\addplot[Set1-7-3, thick, solid, mark=diamond*, mark options={scale=1.3}] table[x index=0, y index = 1] {figs/data/PER_MC_SF11_F20_Fpr8_Pi-3.00_lambda0.00_Coh0_Estim0_sigmaPhi0.00_OS10_phOffset1_CFOJitter0_JittPerc0.00_2Eqs1_0.dat};
		\label{SF11ferMCnoncoh}

		\addplot[black, thick, dotted, mark=star, mark options={scale=0.6, solid}] table[x index=0, y index = 1] {figs/data/PER_APP_FER_NONcoh_SF7_F20_Pi-3.00_tauCFO0.00_OS5_newCorrected_window5_s2_comb8_noOmega.dat};
		\label{SF7ferAPPnoncoh}
		\addplot[black, thick, dotted, mark=pentagon*, mark options={scale=0.6, solid}] table[x index=0, y index = 1] {figs/data/APP_FER_NONcoh_SF9_F20_Pi-3.00_tauCFO0.00_OS5_newCorrected_window5_noOmega_IoT.dat};
		\label{SF9ferAPPnoncoh}
		\addplot[black, thick, dotted, mark=diamond*, mark options={scale=0.6, solid}] table[x index=0, y index = 1] {figs/data/APP_FER_NONcoh_SF11_F20_Pi-3.00_OS5.dat};
		\label{SF11ferAPPnoncoh}


		\addplot[Set1-7-5, thick, solid, mark=*, mark options={scale=1.2}] table[x index=0, y index = 1] {figs/data/PER_MC_SF7_F20_Fpr8_Pi-3.00_lambda0.00_Coh1_Estim0_sigmaPhi0.00_OS10_phOffset1_CFOJitter0_JittPerc0.00_2Eqs1_0.dat};
		\label{SF7ferMCcoh}
		\addplot[Set1-7-2, thick, solid, mark=square*, mark options={scale=1.05}] table[x index=0, y index = 1] {figs/data/PER_MC_SF9_F20_Fpr8_Pi-3.00_lambda0.00_Coh1_Estim0_sigmaPhi0.00_OS10_phOffset1_CFOJitter0_JittPerc0.00_2Eqs1_0.dat};
		\label{SF9ferMCcoh}
		\addplot[Set1-7-3, thick, solid, mark=triangle*, mark options={scale=1.3}] table[x index=0, y index = 1] {figs/data/PER_MC_SF11_F20_Fpr8_Pi-3.00_lambda0.00_Coh1_Estim0_sigmaPhi0.00_OS10_phOffset1_CFOJitter0_JittPerc0.00_2Eqs1_0.dat};
		\label{SF11ferMCcoh}

		\addplot[black, thick, dotted, mark=*, mark options={scale=0.6, solid}] table[x index=0, y index = 1] {figs/data/PER_APP_FER_coh_SF7_F20_Pi-3.00_tauCFO0.00_OS5_newCorrected_window5_s2_comb8.dat};
		\label{SF7ferAPPcoh}
		\addplot[black, thick, dotted, mark=square*, mark options={scale=0.6, solid}] table[x index=0, y index = 1] {figs/data/APP_FER_coh_SF9_F20_Pi-3.00_tauCFO0.00_OS5_newCorrected_window5_s2_setN8_IoT.dat};
		\label{SF9ferAPPcoh}
		\addplot[black, thick, dotted, mark=triangle*, mark options={scale=0.6, solid}] table[x index=0, y index = 1] {figs/data/APP_FER_coh_SF11_F20_Pi-3.00_OS5.dat};
		\label{SF11ferAPPcoh}

		\node [draw,fill=white,inner sep=2pt] at (rel axis cs: 0.18,0.14) {
		\tiny
		\begin{tabular}{lcc}
			\multicolumn{3}{c}{\scriptsize\underline{Coherent}} \\
								& MC 								& Approx. \\
			SF${=}7$:	 	& \ref{SF7ferMCcoh} 	&	\ref{SF7ferAPPcoh} \\
			SF${=}9$: 	& \ref{SF9ferMCcoh} & \ref{SF9ferAPPcoh} \\
			SF${=}11$: 	& \ref{SF11ferMCcoh} & \ref{SF11ferAPPcoh}
		\end{tabular}};

		\node [draw,fill=white,inner sep=2pt] at (rel axis cs: 0.83,0.85) {
		\tiny
		\begin{tabular}{lcc}
			\multicolumn{3}{c}{\scriptsize\underline{Non-coherent} } \\
								& MC 								& Approx. \\
			SF${=}7$:	 	& \ref{SF7ferMCnoncoh} 	&	\ref{SF7ferAPPnoncoh} \\
			SF${=}9$: 	& \ref{SF9ferMCnoncoh} & \ref{SF9ferAPPnoncoh} \\
			SF${=}11$: 	& \ref{SF11ferMCnoncoh} & \ref{SF11ferAPPnoncoh}
		\end{tabular}};

	\end{semilogyaxis}

\end{tikzpicture}%
	\caption{Frame error rate of the coherent and non-coherent receiver for a packet of length $F=20$ LoRa symbols under AWGN and same-SF interference for $\text{SF}
		\in \left\{7,9,11\right\}$ and $P_I = {-}3$ dB. The approximations for the coherent and the non-coherent case are shown with black dotted lines.}
	\label{fig:ferint}
\end{figure}

In Fig.~\ref{fig:ferint}, we show the results of a Monte Carlo simulation for the FER of both a coherent and a non-coherent LoRa receiver with $\text{SF} \in \left\{7,9,11\right\}$, as well as the corresponding approximation described in Section~\ref{sec:FER_interf}. The frame length is $F=20$ LoRa symbols and the SIR is 3~dB. We observe the same performance difference between the coherent and non-coherent receivers as in the SER curves. Furthermore, we see that the low-complexity approximation for the FER under same-SF interference, described in Section~\ref{sec:FER_interf}, is also very accurate.

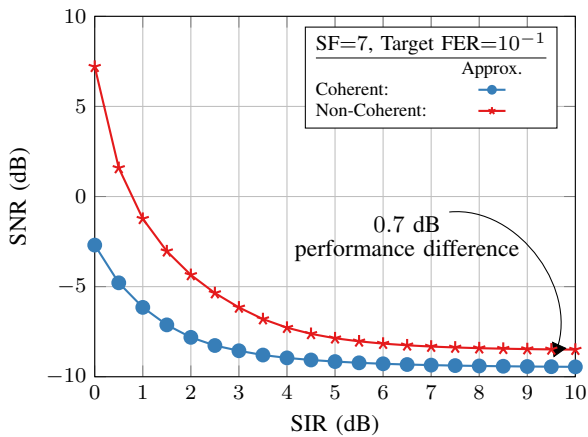
\begin{figure}[t]
	\centering
	\begin{tikzpicture}

	\small

	\begin{axis}[
		width = \figurewidth\columnwidth,
		height = \figureheight\columnwidth,
		xlabel = {SIR (dB)},
		ylabel = {SNR (dB)},
		label style={font=\small},
    tick label style={font=\footnotesize},
		xtick distance=1,
		ytick distance=5,
		ylabel near ticks,
		xlabel near ticks,
		xmin = 0, xmax = 10,
		ymin = -10, ymax = 10,
		grid = both,
		legend image post style={scale=0.6},
	]
		\addplot[Set1-7-2, thick, solid, mark=*, mark options={scale=1.2}] table[x index=0, y index = 1] {figs/data/APP_coh_SNR_SIR_SF7_F20_OS5_PERtarget0.10.dat};
		\label{SF7APPcoh_new}

		\addplot[Set1-7-1, thick, solid, mark=star, mark options={scale=1.2, solid}] table[x index=0, y index = 1] {figs/data/APP_NONcoh_SNR_SIR_SF7_FrLen20_OS5_PERtarget0.10_newCorrected_window5_noOmega.dat};
		\label{SF7APPnoncoh_new}


		\node [draw,fill=white,inner sep=2pt] at (rel axis cs: 0.70,0.83) {
		\scriptsize
		\begin{tabular}{lcc}
			\multicolumn{2}{c}{\footnotesize\underline{SF${=}7$, Target FER${=}10^{-1}$} } \\
						& Approx. \\
		    Coherent:	 	&	\ref{SF7APPcoh_new} \\
			Non-Coherent:	 	&	\ref{SF7APPnoncoh_new} 
			
		\end{tabular}};
\node (source) at (axis cs:6.5, -1.5){$0.7$~dB};
\node at (axis cs:6.5, -3){performance difference};
\node (destination) at (axis cs:9.5, -9.4){};
\draw[->,-triangle 60] (source) to[bend right=-60] (destination);

	\end{axis}

\end{tikzpicture}%
	\caption{Required SNR for a target frame error rate of $10^{-1}$ as a function of the SIR for a packet length of $F=20$ LoRa symbols for $\text{SF}=7$ for coherent and non-coherent receiver.}
	\label{fig:SNR_SIR_FERtarget}
\end{figure}

In Fig.~\ref{fig:SNR_SIR_FERtarget}, we show the required SNR for $\text{SF} = 7$ with a target FER performance of $10^{-1}$, for different SIR levels. We show results for both the coherent and the non-coherent receiver for frames of $20$ LoRa symbols. As expected, for both receivers, there is an increase in the required SNR to obtain the same FER, as the interference power increases. For the chosen target FER of $10^{-1}$ we can observe that the coherent receiver requires a much lower increase in the required SNR compared to the non-coherent receiver. We can also clearly observe that the $~0.7$~dB performance difference between the coherent and non-coherent receiver reported in~\cite{Elshabrawy2019b,Marquet2019,Baruffa2019} for the AWGN case (i.e., for large SIR) is only the smallest possible performance difference. In fact, the coherent receiver is of increasingly greater importance for increasing levels of interference. 

\subsection{Phase estimation errors}

For coherent detection, a receiver needs to estimate the phase of the received symbols. Although the initial estimation of the phase is easy to do in a LoRa packet, thanks to the preamble, the continuous phase tracking, needed due to the phase drift, cannot be perfect. The phase estimation error due to the imperfect tracking can be modeled as a Gaussian process with variance $\sigma_{\text{tr}}^{2}$, which depends on the phase-drift levels and the quality of the tracking algorithm. 

\begin{figure}
	\centering
	\begin{tikzpicture}

	\small

	\begin{semilogyaxis}[
		width = \figurewidth\columnwidth,
		height = \figureheight\columnwidth,
		xlabel = {SNR (dB)},
		ylabel = {Frame Error Rate},
		label style={font=\small},
    tick label style={font=\footnotesize},
		ylabel near ticks,
		xlabel near ticks,
		xmin = -15, xmax = 3,
		ymin = 1e-4, ymax = 1,
		grid = both,
		legend image post style={scale=0.6},
	]


		\addplot[black, ultra thick, solid, opacity=0.25] table[x index=0, y index = 1] {figs/data/PER_MC_SF7_F20_Fpr8_Pi-3.00_lambda0.00_Coh0_Estim0_sigmaPhi0.00_OS10_phOffset1_CFOJitter0_JittPerc0.00_2Eqs1_0.dat};
		\label{SF7MCnoncoh}


		\addplot[Set1-7-5, thick, solid, mark=*, mark options={scale=1.2}] table[x index=0, y index = 1] {figs/data/PER_MC_SF7_F20_Fpr8_Pi-3.00_lambda0.00_Coh1_Estim0_sigmaPhi0.00_OS10_phOffset1_CFOJitter0_JittPerc0.00_2Eqs1_0.dat};
		\label{SF7MCcoh}
		\addplot[Set1-7-4, thick, solid, mark=*, mark options={scale=1.2}] table[x index=0, y index = 1] {figs/data/PER_MC_SF7_F20_Fpr8_Pi-3.00_lambda0.00_Coh1_Estim2_sigmaPhi0.20_OS10_phOffset1_CFOJitter0_JittPerc0.00_2Eqs1_0.dat};
		\label{SF7MCcoh_sigma02}
        \addplot[Set1-7-1, thick, solid, mark=*, mark options={scale=1.2}] table[x index=0, y index = 1] {figs/data/PER_MC_SF7_F20_Fpr8_Pi-3.00_lambda0.00_Coh1_Estim2_sigmaPhi0.30_OS10_phOffset1_CFOJitter0_JittPerc0.00_2Eqs1_0.dat};
        \label{SF7MCcoh_sigma03}
		\addplot[Set1-7-3, thick, solid, mark=*, mark options={scale=1.2}] table[x index=0, y index = 1] {figs/data/PER_MC_SF7_F20_Fpr8_Pi-3.00_lambda0.00_Coh1_Estim2_sigmaPhi0.40_OS10_phOffset1_CFOJitter0_JittPerc0.10_2Eqs1_0.dat};
		\label{SF7MCcoh_sigma04}

		\node [draw,fill=white,inner sep=2pt] at (rel axis cs: 0.19,0.2) {
		\tiny
		\begin{tabular}{lcc}
			\multicolumn{3}{c}{\scriptsize\underline{Coherent, SF${=}7$}} \\
								& MC 			\\				
			Perfect estim.:	 	& \ref{SF7MCcoh} \\
			$\sigma^{2}{=}0.2$: 	& \ref{SF7MCcoh_sigma02} \\
			$\sigma^{2}{=}0.3$: 	& \ref{SF7MCcoh_sigma03} \\
			$\sigma^{2}{=}0.4$:     & \ref{SF7MCcoh_sigma04}
		\end{tabular}};


	\end{semilogyaxis}

\end{tikzpicture}%
	\caption{Frame error rate of the coherent receiver under AWGN and same-SF interference with perfect estimation, and for estimation error $\sigma^{2}
		\in \left\{0.2,0.3,0.4\right\}$. The packet length is $F=20$ LoRa symbols, $\text{SF}{=}7$ and $P_I = {-}3$ dB. The performance of the non-coherent receiver is shown in thick transparent line.}
	\label{fig:ferint_estimError}
\end{figure}
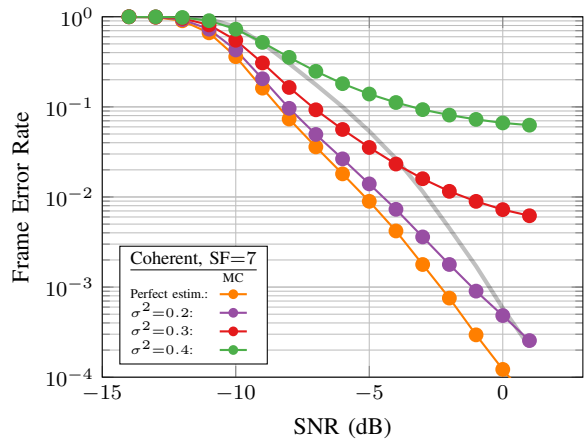

In Fig.~\ref{fig:ferint_estimError}, we show the performance degradation of a coherent LoRa receiver for three different estimation error variances, $\sigma^{2} \in \left\{0.2,0.3,0.4\right\}$, as well as for perfect phase estimation. The frame length is $F=20$ LoRa symbols, the spreading factor is $\text{SF}=7$, and $\text{SIR}=3$~dB. The results are compared to the non-coherent receiver, shown in a thick gray line. We observe that the tracking phase-estimation error plays a significant role in the performance of a coherent receiver. With small estimation errors, the coherent receiver can still be competitive to the non-coherent receiver. However, for a coherent receiver to be worth applying compared to a non-coherent receiver, a careful design of the phase-tracking algorithm is needed. 

\section{Conclusion} \label{sec:conclusion}
In this work we model and investigate the performance of coherent LoRa receivers under same-SF interference. We derive very accurate error rate expressions for the symbol error rate and the frame error rate and we show that the performance improvement of $0.7$~dB reported previously for the coherent receiver compared to the non-coherent receiver in the AWGN scenario, is only the smallest possible benefit. The benefit of employing a coherent LoRa receiver increases for higher levels of interference and can reach values close to $10$~dB for an SIR of 0~dB. This observation renders coherent LoRa receivers particularly attractive for congested LoRa networks with a high probability of same-SF packet collisions.

\section*{Acknowledgment}
The authors thank Mathieu Xhonneux for helpful discussions on the LoRa signal model in presence of both sampling time offsets and carrier frequency offsets. 

\appendix[]
The terms $A_{k,j}$ and $\theta_{k,j}$, for $j \in\{1,\dots,4\}$, in~\eqref{eq:ReRk} are
{\small
\begin{align}
A_{k,1} & {=} \frac{\sin \left( \frac{\pi}{N} (s_{I_{1}}{-}k{-}\tau {+} \tau_{\text{cfo}})\max\left( \lceil \tau \rceil{-}s_{I_{1}},0\right) \right)}{\sin \left( \frac{\pi}{N} (s_{I_{1}}{-}k{-}\tau+\tau_{\text{cfo}}) \right)}, \label{eq:A1} \\
A_{k,2} & {=} \frac{\sin \left( \frac{\pi}{N} (s_{I_{1}}{-}k{-}\tau {+} \tau_{\text{cfo}})\left(\lceil \tau\rceil{-} \max\left( \lceil \tau \rceil{-}s_{I_{1}} ,0\right) \right) \right)}{\sin \left( \frac{\pi}{N} (s_{I_{1}}{-}k{-}\tau+\tau_{\text{cfo}}) \right)}, \label{eq:A2} \\
A_{k,3} &=  \frac{\sin \left( \frac{\pi}{N} (s_{I_{2}}{-}k{-}\tau{+}\tau_{\text{cfo}})\left(\min\left(N{-}s_{I_{2}}{+}\lceil \tau\rceil,N\right) {-} \lceil \tau\rceil\right)\right)}{\sin \left( \frac{\pi}{N} (s_{I_{2}}{-}k{-}\tau{+}\tau_{\text{cfo}}) \right)}, \label{eq:A3}\\
A_{k,4} &=  \frac{\sin \left( \frac{\pi}{N} (s_{I_{2}}{-}k{-}\tau{+}\tau_{\text{cfo}})\max\left(s_{I_{2}}{-}\lceil \tau \rceil,0\right)\right)}{\sin \left( \frac{\pi}{N} (s_{I_{2}}{-}k{-}\tau{+}\tau_{\text{cfo}}) \right)}, \label{eq:A4}
\end{align}
}
and
{\small
\begin{align}
\theta_{k,1} &= \frac{\pi}{N} \bigg( {-}\tau^2 {+} \tau N {+} s_{I_{1}}(2\tau{-} \lceil \tau \rceil{+}s_{I_{1}} {+}1) {+}  \nonumber\\
&{+}  \left(k{+}\tau{-}\tau_{\text{cfo}}\right)(\lceil \tau \rceil{-}s_{I_{1}} {-}1) {-}  2N\left(m{-}1\right)\tau_{\text{cfo}} {-}\omega\frac{N}{\pi} \bigg),\\
\theta_{k,2} & = \frac{\pi}{N} \bigg( {-}\tau^2 {-} \tau N {+} s_{I_{1}}(2\tau{-}\lceil \tau\rceil{+} \max\left( \lceil \tau \rceil{-}s_{I_{1}} ,0\right) {+}1) {+} \nonumber\\
&  {+} \left(k{+}\tau{-}\tau_{\text{cfo}}\right)(\lceil \tau\rceil{+} \max\left( \lceil \tau \rceil{-}s_{I_{1}} ,0\right) {-}1) {+} \nonumber\\
&  {-} 2N\left(m{-}1\right)\tau_{\text{cfo}} {-}\omega\frac{N}{\pi} \bigg), \\
\theta_{k,3} & = \frac{\pi}{N} \bigg( {-}\tau^2 {-} \tau N {+} s_{I_{2}}(2\tau{-} \min\left(N{-}s_{I_{2}}{+}\lceil \tau\rceil,N\right) {+} \lceil \tau\rceil{+}1) {+} \nonumber\\
&  {+} \left(k{+}\tau{-}\tau_{\text{cfo}}\right)(\min\left(N{-}s_{I_{2}}{+}\lceil \tau\rceil,N\right) {-} \lceil \tau\rceil {-}1) {+} \nonumber\\
&  {-} 2N\left(m{-}1\right)\tau_{\text{cfo}} {-}\omega\frac{N}{\pi} \bigg), \\
\theta_{k,4} & = \frac{\pi}{N} \bigg( {-}\tau^2 {-} 3\tau N {+} s_{I_{2}}(2\tau{-} 2N{+}s_{I_{2}}{-}\lceil \tau\rceil {+}1) {+} \nonumber\\
&  {+} \left(k{+}\tau{-}\tau_{\text{cfo}}\right)(2N{-} s_{I_{2}}{+}\lceil \tau\rceil {-}1) {-} 2N\left(m{-}1\right)\tau_{\text{cfo}} {-}\omega\frac{N}{\pi} \bigg).
\end{align}
}

\bibliographystyle{IEEEtran}
\bibliography{IEEEabrv,refs}

\begin{thebibliography}{10}
\providecommand{\url}[1]{#1}
\csname url@samestyle\endcsname
\providecommand{\newblock}{\relax}
\providecommand{\bibinfo}[2]{#2}
\providecommand{\BIBentrySTDinterwordspacing}{\spaceskip=0pt\relax}
\providecommand{\BIBentryALTinterwordstretchfactor}{4}
\providecommand{\BIBentryALTinterwordspacing}{\spaceskip=\fontdimen2\font plus
\BIBentryALTinterwordstretchfactor\fontdimen3\font minus
  \fontdimen4\font\relax}
\providecommand{\BIBforeignlanguage}[2]{{%
\expandafter\ifx\csname l@#1\endcsname\relax
\typeout{** WARNING: IEEEtran.bst: No hyphenation pattern has been}%
\typeout{** loaded for the language `#1'. Using the pattern for}%
\typeout{** the default language instead.}%
\else
\language=\csname l@#1\endcsname
\fi
#2}}
\providecommand{\BIBdecl}{\relax}
\BIBdecl

\bibitem{Raza2017}
U.~{Raza}, P.~{Kulkarni}, and M.~{Sooriyabandara}, ``Low power wide area
  networks: An overview,'' \emph{IEEE Communications Surveys {\&} Tutorials},
  vol.~19, no.~2, pp. 855--873, Secondquarter 2017.

\bibitem{Haxhibeqiri2018}
J.~Haxhibeqiri, E.~De~Poorter, I.~Moerman, and J.~Hoebeke, ``A survey of
  {LoRaWAN} for {IoT}: From technology to application,'' \emph{Sensors},
  vol.~18, no.~11, 2018.

\bibitem{Boulogeorgos2016}
A.~Boulogeorgos, P.~Diamantoulakis, and G.~Karagiannidis, ``Low power wide area
  networks ({LPWANS}) for {Internet} of {Things} ({IoT}) applications: Research
  challenges and future trends,'' \emph{ArXiv e-prints}, Nov. 2016,
  \url{https://arxiv.org/abs/1611.07449}.

\bibitem{Seller2016}
O.~B. Seller and N.~Sornin, ``Low power long range transmitter,'' US Patent
  9,252,834, Feb., 2016.

\bibitem{Vangelista2017}
L.~{Vangelista}, ``Frequency shift chirp modulation: The {LoRa} modulation,''
  \emph{IEEE Signal Processing Letters}, vol.~24, no.~12, pp. 1818--1821, Dec.
  2017.

\bibitem{Georgiou2017}
O.~{Georgiou} and U.~{Raza}, ``Low power wide area network analysis: Can {LoRa}
  scale?'' \emph{IEEE Wireless Communications Letters}, vol.~6, no.~2, pp.
  162--165, Apr. 2017.

\bibitem{Bor2016}
M.~C. Bor, U.~Roedig, T.~Voigt, and J.~M. Alonso, ``Do {LoRa} low-power
  wide-area networks scale?'' in \emph{ACM International Conference on
  Modeling, Analysis and Simulation of Wireless and Mobile Systems}, ser. MSWiM
  '16.\hskip 1em plus 0.5em minus 0.4em\relax New York, NY, USA: ACM, 2016, pp.
  59--67.

\bibitem{BenTemim2020}
M.~A. {Ben Temim}, G.~{Ferré}, B.~{Laporte-Fauret}, D.~{Dallet}, B.~{Minger},
  and L.~{Fuché}, ``An enhanced receiver to decode superposed {LoRa}-like
  signals,'' \emph{IEEE Internet of Things Journal}, vol.~7, no.~8, pp.
  7419--7431, Aug. 2020.

\bibitem{Orfanidis2017}
C.~{Orfanidis}, L.~M. {Feeney}, M.~{Jacobsson}, and P.~{Gunningberg},
  ``Investigating interference between {LoRa} and {IEEE} 802.15.4g networks,''
  in \emph{IEEE International Conference on Wireless and Mobile Computing,
  Networking and Communications (WiMob)}, Oct. 2017, pp. 1--8.

\bibitem{Marquez2020}
L.~E. {Marquez}, A.~{Osorio}, M.~{Calle}, J.~C. {Velez}, A.~{Serrano}, and
  J.~E. {Candelo-Becerra}, ``On the use of {LoRaWAN} in smart cities: A study
  with blocking interference,'' \emph{IEEE Internet of Things Journal}, vol.~7,
  no.~4, pp. 2806--2815, Apr. 2020.

\bibitem{Voigt2016}
T.~Voigt, M.~Bor, U.~Roedig, and J.~Alonso, ``Mitigating inter-network
  interference in {LoRa} networks,'' in \emph{Proceedings of the 2017
  International Conference on Embedded Wireless Systems and Networks {(EWSN)}},
  Feb. 2017.

\bibitem{Haxhibeqiri2017b}
J.~Haxhibeqiri, F.~Van~den Abeele, I.~Moerman, and J.~Hoebeke, ``{LoRa}
  scalability: A simulation model based on interference measurements,''
  \emph{Sensors}, vol.~17, no.~6, Jun. 2017.

\bibitem{Ferrari2017}
P.~{Ferrari}, A.~{Flammini}, M.~{Rizzi}, E.~{Sisinni}, and M.~{Gidlund}, ``On
  the evaluation of {LoRaWAN} virtual channels orthogonality for dense
  distributed systems,'' in \emph{IEEE International Workshop on Measurement
  and Networking (M{\&}N)}, Sep. 2017, pp. 1--6.

\bibitem{Fernandes2019}
R.~{Fernandes}, R.~{Oliveira}, M.~{Lu\'is}, and S.~{Sargento}, ``On the real
  capacity of {LoRa} networks: the impact of non-destructive communications,''
  \emph{IEEE Communications Letters}, vol.~23, no.~12, pp. 2437--2441, Dec.
  2019.

\bibitem{Croce2017}
D.~Croce, M.~Gucciardo, I.~Tinnirello, D.~Garlisi, and S.~Mangione, ``Impact of
  spreading factor imperfect orthogonality in {LoRa} communications,'' in
  \emph{Digital Communication. Towards a Smart and Secure Future Internet},
  A.~Piva, I.~Tinnirello, and S.~Morosi, Eds.\hskip 1em plus 0.5em minus
  0.4em\relax Cham: Springer International Publishing, 2017, pp. 165--179.

\bibitem{Croce2018}
D.~{Croce}, M.~{Gucciardo}, S.~{Mangione}, G.~{Santaromita}, and
  I.~{Tinnirello}, ``Impact of {LoRa} imperfect orthogonality: Analysis of
  link-level performance,'' \emph{IEEE Communications Letters}, vol.~22, no.~4,
  pp. 796--799, Apr. 2018.

\bibitem{Goursaud2015}
C.~Goursaud and J.-M. Gorce, ``{Dedicated networks for {IoT} : {PHY} / {MAC}
  state of the art and challenges},'' \emph{{EAI endorsed trans. on Internet of
  Things}}, Oct. 2015.

\bibitem{Mikhaylov2017}
K.~{Mikhaylov}, J.~{Pet\"{a}j\"{a}j\"{a}rvi}, and J.~{Janhunen}, ``On {LoRaWAN}
  scalability: Empirical evaluation of susceptibility to inter-network
  interference,'' in \emph{2017 European Conference on Networks and
  Communications (EuCNC)}, Jun. 2017.

\bibitem{Feltrin2018}
L.~{Feltrin}, C.~{Buratti}, E.~{Vinciarelli}, R.~{De Bonis}, and R.~{Verdone},
  ``{LoRaWAN}: Evaluation of link- and system-level performance,'' \emph{IEEE
  Internet of Things Journal}, vol.~5, no.~3, pp. 2249--2258, Jun. 2018.

\bibitem{Elshabrawy2018b}
T.~{Elshabrawy} and J.~{Robert}, ``Analysis of {BER} and coverage performance
  of {LoRa} modulation under same spreading factor interference,'' in
  \emph{IEEE International Symposium on Personal, Indoor and Mobile Radio
  Communications (PIMRC)}, Sep. 2018.

\bibitem{Elshabrawy2019}
T.~{Elshabrawy} and J.~{Robert}, ``Capacity planning of {LoRa} networks with
  joint noise-limited and interference-limited coverage considerations,''
  \emph{IEEE Sensors Journal}, pp. 1--1, Feb. 2019.

\bibitem{Mahmood2019}
A.~{Mahmood}, E.~{Sisinni}, L.~{Guntupalli}, R.~{Rondón}, S.~A. {Hassan}, and
  M.~{Gidlund}, ``Scalability analysis of a {LoRa} network under imperfect
  orthogonality,'' \emph{IEEE Transactions on Industrial Informatics}, vol.~15,
  no.~3, pp. 1425--1436, Mar. 2019.

\bibitem{Waret2019}
A.~{Waret}, M.~{Kaneko}, A.~{Guitton}, and N.~{El Rachkidy}, ``{LoRa}
  throughput analysis with imperfect spreading factor orthogonality,''
  \emph{IEEE Wireless Communications Letters}, vol.~8, no.~2, pp. 408--411,
  Apr. 2019.

\bibitem{Croce2020}
D.~{Croce}, M.~{Gucciardo}, S.~{Mangione}, G.~{Santaromita}, and
  I.~{Tinnirello}, ``{LoRa} technology demystified: From link behavior to
  cell-level performance,'' \emph{IEEE Transactions on Wireless
  Communications}, vol.~19, no.~2, pp. 822--834, Feb. 2020.

\bibitem{Georgiou2020}
O.~{Georgiou}, C.~{Psomas}, and I.~{Krikidis}, ``Coverage scalability analysis
  of multi-cell {LoRa} networks,'' in \emph{IEEE International Conference on
  Communications (ICC)}, Jun. 2020, pp. 1--7.

\bibitem{Pop2017}
A.~{Pop}, U.~{Raza}, P.~{Kulkarni}, and M.~{Sooriyabandara}, ``Does
  bidirectional traffic do more harm than good in {LoRaWAN} based {LPWA}
  networks?'' in \emph{IEEE Global Communications Conference (GLOBECOM)}, Dec.
  2017, pp. 1--6.

\bibitem{Abdelfadeel2020}
K.~Q. {Abdelfadeel}, D.~{Zorbas}, V.~{Cionca}, and D.~{Pesch}, ``{$FREE$} --
  fine-grained scheduling for reliable and energy-efficient data collection in
  {LoRaWAN},'' \emph{IEEE Internet of Things Journal}, vol.~7, no.~1, pp.
  669--683, Jan. 2020.

\bibitem{Callebaut2019}
G.~{Callebaut} and L.~{Van der Perre}, ``Characterization of {LoRa}
  point-to-point path-loss: Measurement campaigns and modeling considering
  censored data,'' \emph{IEEE Internet of Things Journal}, vol.~7, no.~3, pp.
  1910--1918, Mar. 2020.

\bibitem{Abeele2017}
F.~{Van den Abeele}, J.~{Haxhibeqiri}, I.~{Moerman}, and J.~{Hoebeke},
  ``Scalability analysis of large-scale {LoRaWAN} networks in ns-3,''
  \emph{IEEE Internet of Things Journal}, vol.~4, no.~6, pp. 2186--2198, Dec.
  2017.

\bibitem{Reynders2018}
B.~Reynders, Q.~Wang, and S.~Pollin, ``A {LoRaWAN} module for {ns-3}:
  Implementation and evaluation,'' in \emph{Workshop on ns-3}, ser. WNS3
  '18.\hskip 1em plus 0.5em minus 0.4em\relax New York, NY, USA: ACM, 2018, pp.
  61--68.

\bibitem{Reynders2018b}
B.~{Reynders}, Q.~{Wang}, P.~{Tuset-Peiro}, X.~{Vilajosana}, and S.~{Pollin},
  ``Improving reliability and scalability of {LoRaWANs} through lightweight
  scheduling,'' \emph{IEEE Internet of Things Journal}, vol.~5, no.~3, pp.
  1830--1842, Jun. 2018.

\bibitem{Magrin2020}
D.~{Magrin}, M.~{Capuzzo}, and A.~{Zanella}, ``A thorough study of {LoRaWAN}
  performance under different parameter settings,'' \emph{IEEE Internet of
  Things Journal}, vol.~7, no.~1, pp. 116--127, Jan. 2020.

\bibitem{To2018}
T.~{To} and A.~{Duda}, ``Simulation of {LoRa} in {NS-3}: Improving {LoRa}
  performance with {CSMA},'' in \emph{2018 IEEE International Conference on
  Communications (ICC)}, May 2018, pp. 1--7.

\bibitem{Kouvelas2018}
N.~Kouvelas, V.~Rao, and R.~R.~V. Prasad, ``Employing p-{CSMA} on a {LoRa}
  network simulator,'' May 2018.

\bibitem{Slabicki2018}
M.~{Slabicki}, G.~{Premsankar}, and M.~{Di Francesco}, ``Adaptive configuration
  of {LoRa} networks for dense {IoT} deployments,'' in \emph{NOMS 2018 - 2018
  IEEE/IFIP Network Operations and Management Symposium}, Apr. 2018, pp. 1--9.

\bibitem{Yousuf2018}
A.~M. {Yousuf}, E.~M. {Rochester}, B.~{Ousat}, and M.~{Ghaderi}, ``Throughput,
  coverage and scalability of {LoRa} {LPWAN} for {Internet} of {Things},'' in
  \emph{2018 IEEE/ACM 26th International Symposium on Quality of Service
  (IWQoS)}, Jun. 2018, pp. 1--10.

\bibitem{Centenaro2017}
M.~{Centenaro}, L.~{Vangelista}, and R.~{Kohno}, ``On the impact of downlink
  feedback on {LoRa} performance,'' in \emph{IEEE Annual International
  Symposium on Personal, Indoor, and Mobile Radio Communications (PIMRC)}, Oct.
  2017, pp. 1--6.

\bibitem{Finnegan2020}
J.~{Finnegan}, R.~{Farrell}, and S.~{Brown}, ``Analysis and enhancement of the
  {LoRaWAN} adaptive data rate scheme,'' \emph{IEEE Internet of Things
  Journal}, vol.~7, no.~8, pp. 7171--7180, Aug. 2020.

\bibitem{Furtado2020}
A.~{Furtado}, J.~{Pacheco}, and R.~{Oliveira}, ``{PHY/MAC} uplink performance
  of {LoRa} class {A} networks,'' \emph{IEEE Internet of Things Journal},
  vol.~7, no.~7, pp. 6528--6538, Jul. 2020.

\bibitem{Reynders2016}
B.~{Reynders}, W.~{Meert}, and S.~{Pollin}, ``Range and coexistence analysis of
  long range unlicensed communication,'' in \emph{International Conference on
  Telecommunications (ICT)}, May 2016, pp. 1--6.

\bibitem{Reynders2016b}
B.~{Reynders} and S.~{Pollin}, ``Chirp spread spectrum as a modulation
  technique for long range communication,'' in \emph{2016 Symposium on
  Communications and Vehicular Technologies (SCVT)}, Nov. 2016.

\bibitem{Elshabrawy2018}
T.~Elshabrawy and J.~Robert, ``Closed-form approximation of {LoRa} modulation
  {BER} performance,'' \emph{IEEE Communications Letters}, vol.~22, no.~9, pp.
  1778--1781, Sep. 2018.

\bibitem{FerreiraDias_2019}
C.~Ferreira~Dias, E.~Rodrigues~de Lima, and G.~Fraidenraich, ``Bit error rate
  closed-form expressions for {LoRa} systems under {Nakagami} and {Rice} fading
  channels,'' \emph{Sensors}, vol.~19, Oct 2019.

\bibitem{Baruffa2019}
G.~{Baruffa}, L.~{Rugini}, V.~{Mecarelli}, L.~{Germani}, and F.~{Frescura},
  ``Coded {LoRa} performance in wireless channels,'' in \emph{IEEE Annual
  International Symposium on Personal, Indoor and Mobile Radio Communications
  (PIMRC)}, Sep. 2019, pp. 1--6.

\bibitem{Courjault2019}
\BIBentryALTinterwordspacing
J.~Courjault, B.~Vrigneau, M.~Gautier, and O.~Berder, ``{Accurate {LoRa}
  Performance evaluation using Marcum function},'' in \emph{{Globecom 2019}},
  Dec. 2019. [Online]. Available:
  \url{https://hal.archives-ouvertes.fr/hal-02418665}
\BIBentrySTDinterwordspacing

\bibitem{Afisiadis2019c}
O.~Afisiadis, A.~Burg, and A.~Balatsoukas-Stimming, ``Coded {LoRa} frame error
  rate analysis,'' in \emph{IEEE International Conference on Communications
  (ICC)}, Jun. 2020.

\bibitem{Baruffa2020}
G.~{Baruffa}, L.~{Rugini}, L.~{Germani}, and F.~{Frescura},
  ``\BIBforeignlanguage{en}{Error probability performance of chirp modulation
  in uncoded and coded {LoRa} systems},''
  \emph{\BIBforeignlanguage{en}{Elsevier {Digital} {Signal} {Processing}}},
  vol. 106, Nov. 2020.

\bibitem{Guo2020}
Y.~{Guo} and Z.~{Liu}, ``Time-delay-estimation-liked detection algorithm for
  {LoRa} signals over multipath channels,'' \emph{IEEE Wireless Communications
  Letters}, vol.~9, no.~7, pp. 1093--1096, Jul. 2020.

\bibitem{Xu2020}
W.~{Xu}, J.~Y. {Kim}, W.~{Huang}, S.~S. {Kanhere}, S.~K. {Jha}, and W.~{Hu},
  ``Measurement, characterization, and modeling of {LoRa} technology in
  multifloor buildings,'' \emph{IEEE Internet of Things Journal}, vol.~7,
  no.~1, pp. 298--310, Jan. 2020.

\bibitem{Afisiadis2020}
O.~{Afisiadis}, M.~{Cotting}, A.~{Burg}, and A.~{Balatsoukas-Stimming}, ``On
  the error rate of the {LoRa} modulation with interference,'' \emph{IEEE
  Transactions on Wireless Communications}, vol.~19, no.~2, pp. 1292--1304,
  Feb. 2020.

\bibitem{Elshabrawy2019b}
T.~{Elshabrawy}, P.~{Edward}, M.~{Ashour}, and J.~{Robert}, ``On the different
  mathematical realizations for the digital synthesis of {LoRa}-based
  modulation,'' in \emph{European Wireless 2019; 25th European Wireless
  Conference}, May 2019, pp. 1--6.

\bibitem{Marquet2019}
A.~{Marquet}, N.~{Montavont}, and G.~Z. {Papadopoulos}, ``Investigating
  theoretical performance and demodulation techniques for {LoRa},'' in
  \emph{IEEE International Symposium on "A World of Wireless, Mobile and
  Multimedia Networks" (WoWMoM)}, June 2019, pp. 1--6.

\bibitem{Marquet2020}
A.~{Marquet}, N.~{Montavont}, and G.~Z. {Papadopoulos}, ``Towards an {SDR}
  implementation of {LoRa}: Reverse-engineering, demodulation strategies and
  assessment over {Rayleigh} channel,'' \emph{Computer Communications}, vol.
  153, pp. 595 -- 605, 2020.

\bibitem{Elshabrawy2019c}
T.~{Elshabrawy} and J.~{Robert}, ``Interleaved chirp spreading {LoRa}-based
  modulation,'' \emph{IEEE {Internet} of {Things} Journal}, vol.~6, no.~2, pp.
  3855--3863, Apr. 2019.

\bibitem{Ghanaatian2019}
R.~{Ghanaatian}, O.~{Afisiadis}, M.~{Cotting}, and A.~{Burg}, ``{LoRa} digital
  receiver analysis and implementation,'' in \emph{IEEE International
  Conference on Acoustics, Speech and Signal Processing (ICASSP)}, May 2019.

\bibitem{Chiani2019}
M.~{Chiani} and A.~{Elzanaty}, ``On the {LoRa} modulation for {IoT}: Waveform
  properties and spectral analysis,'' \emph{IEEE Internet of Things Journal},
  vol.~6, no.~5, pp. 8463--8470, Oct. 2019.

\bibitem{Karagiannidis2007}
G.~K. {Karagiannidis} and A.~S. {Lioumpas}, ``An improved approximation for the
  {Gaussian} {Q}-function,'' \emph{IEEE Communications Letters}, vol.~11,
  no.~8, pp. 644--646, Aug. 2007.

\bibitem{Bernier2019}
C.~{Bernier}, F.~{Dehmas}, and N.~{Deparis}, ``Low complexity {LoRa} frame
  synchronization for ultra-low power software-defined radios,'' \emph{IEEE
  Transactions on Communications}, vol.~68, no.~5, pp. 3140--3152, May 2020.

\bibitem{Xhonneux2019}
M.~Xhonneux, D.~Bol, and J.~Louveaux, ``A low-complexity synchronization scheme
  for {LoRa} end nodes,'' \emph{ArXiv e-prints}, Dec. 2019,
  \url{https://arxiv.org/abs/1912.11344}.

\bibitem{Tapparel2020}
J.~{Tapparel}, O.~{Afisiadis}, P.~{Mayoraz}, A.~{Balatsoukas-Stimming}, and
  A.~{Burg}, ``An open-source {LoRa} physical layer prototype on {GNU} radio,''
  in \emph{International Workshop on Signal Processing Advances in Wireless
  Communications (SPAWC)}, May 2020, pp. 1--5.

\end{thebibliography}


\begin{thebibliography}{10}
\providecommand{\url}[1]{#1}
\csname url@samestyle\endcsname
\providecommand{\newblock}{\relax}
\providecommand{\bibinfo}[2]{#2}
\providecommand{\BIBentrySTDinterwordspacing}{\spaceskip=0pt\relax}
\providecommand{\BIBentryALTinterwordstretchfactor}{4}
\providecommand{\BIBentryALTinterwordspacing}{\spaceskip=\fontdimen2\font plus
\BIBentryALTinterwordstretchfactor\fontdimen3\font minus
  \fontdimen4\font\relax}
\providecommand{\BIBforeignlanguage}[2]{{%
\expandafter\ifx\csname l@#1\endcsname\relax
\typeout{** WARNING: IEEEtran.bst: No hyphenation pattern has been}%
\typeout{** loaded for the language `#1'. Using the pattern for}%
\typeout{** the default language instead.}%
\else
\language=\csname l@#1\endcsname
\fi
#2}}
\providecommand{\BIBdecl}{\relax}
\BIBdecl

\bibitem{Seller2016}
O.~B. Seller and N.~Sornin, ``Low power long range transmitter,'' US Patent
  9,252,834, Feb., 2016.

\bibitem{Raza2017}
U.~{Raza}, P.~{Kulkarni}, and M.~{Sooriyabandara}, ``Low power wide area
  networks: An overview,'' \emph{IEEE Commun. Surveys Tutorials}, vol.~19,
  no.~2, pp. 855--873, Secondquarter 2017.

\bibitem{Haxhibeqiri2018}
J.~Haxhibeqiri, E.~De~Poorter, I.~Moerman, and J.~Hoebeke, ``A survey of
  {LoRaWAN} for {IoT}: From technology to application,'' \emph{Sensors},
  vol.~18, no.~11, 2018.

\bibitem{Boulogeorgos2016}
A.~Boulogeorgos, P.~Diamantoulakis, and G.~Karagiannidis, ``Low power wide area
  networks ({LPWANS}) for internet of things ({IoT}) applications: Research
  challenges and future trends,'' \emph{ArXiv e-prints}, Nov. 2016,
  \url{https://arxiv.org/abs/1611.07449}.

\bibitem{Elshabrawy2018}
T.~Elshabrawy and J.~Robert, ``Closed-form approximation of {LoRa} modulation
  {BER} performance,'' \emph{IEEE Commun. Lett.}, vol.~22, no.~9, pp.
  1778--1781, Sep. 2018.

\bibitem{FerreiraDias_2019}
C.~Ferreira~Dias, E.~Rodrigues~de Lima, and G.~Fraidenraich, ``Bit error rate
  closed-form expressions for {LoRa} systems under nakagami and rice fading
  channels,'' \emph{Sensors}, vol.~19, Oct 2019.

\bibitem{Baruffa2019}
G.~{Baruffa}, L.~{Rugini}, V.~{Mecarelli}, L.~{Germani}, and F.~{Frescura},
  ``Coded {LoRa} performance in wireless channels,'' in \emph{2019 IEEE 30th
  Annual International Symposium on Personal, Indoor and Mobile Radio
  Communications (PIMRC)}, Sep. 2019, pp. 1--6.

\bibitem{Courjault2019}
\BIBentryALTinterwordspacing
J.~Courjault, B.~Vrigneau, M.~Gautier, and O.~Berder, ``{Accurate {LoRa}
  Performance evaluation using Marcum function},'' in \emph{{Globecom 2019}},
  Dec. 2019. [Online]. Available:
  \url{https://hal.archives-ouvertes.fr/hal-02418665}
\BIBentrySTDinterwordspacing

\bibitem{Callebaut2019}
G.~{Callebaut} and L.~{Van der Perre}, ``Characterization of {LoRa}
  point-to-point path-loss: Measurement campaigns and modeling considering
  censored data,'' \emph{IEEE Internet of Things Journal}, 2019 (early access).

\bibitem{Elshabrawy2018b}
T.~{Elshabrawy} and J.~{Robert}, ``Analysis of {BER} and coverage performance
  of {LoRa} modulation under same spreading factor interference,'' in
  \emph{IEEE Int. Symp. on Personal, Indoor and Mobile Radio Communications
  (PIMRC)}, Sep. 2018.

\bibitem{Afisiadis2019b}
O.~{Afisiadis}, M.~{Cotting}, A.~{Burg}, and A.~{Balatsoukas-Stimming}, ``On
  the error rate of the {LoRa} modulation with interference,'' \emph{IEEE
  Transactions on Wireless Communications}, Nov. 2019 (early access).

\bibitem{Afisiadis2019c}
O.Afisiadis, A.~Burg, and A.~Balatsoukas-Stimming, ``Coded {LoRa} frame error
  rate analysis,'' \emph{ArXiv e-prints}, Nov. 2019,
  \url{https://arxiv.org/abs/1911.10245}.

\bibitem{Elshabrawy2019c}
T.~{Elshabrawy}, P.~{Edward}, M.~{Ashour}, and J.~{Robert}, ``On the different
  mathematical realizations for the digital synthesis of lora-based
  modulation,'' in \emph{European Wireless 2019; 25th European Wireless
  Conference}, May 2019, pp. 1--6.

\bibitem{Marquet2019}
A.~{Marquet}, N.~{Montavont}, and G.~Z. {Papadopoulos}, ``Investigating
  theoretical performance and demodulation techniques for {LoRa},'' in
  \emph{2019 IEEE 20th International Symposium on "A World of Wireless, Mobile
  and Multimedia Networks" (WoWMoM)}, June 2019, pp. 1--6.

\bibitem{Marquet2020}
A.~{Marquet}, N.~{Montavont}, and G.~Z. {Papadopoulos}, ``Towards an {SDR}
  implementation of {LoRa}: Reverse-engineering, demodulation strategies and
  assessment over {Rayleigh} channel,'' \emph{Computer Communications}, vol.
  153, pp. 595 -- 605, 2020.

\bibitem{Ghanaatian2019}
R.~{Ghanaatian}, O.~{Afisiadis}, M.~{Cotting}, and A.~{Burg}, ``{LoRa} digital
  receiver analysis and implementation,'' in \emph{ICASSP 2019 - 2019 IEEE
  International Conference on Acoustics, Speech and Signal Processing
  (ICASSP)}, May 2019.

\bibitem{Croce2018}
D.~{Croce}, M.~{Gucciardo}, S.~{Mangione}, G.~{Santaromita}, and
  I.~{Tinnirello}, ``Impact of {LoRa} imperfect orthogonality: Analysis of
  link-level performance,'' \emph{IEEE Commun. Lett.}, vol.~22, no.~4, pp.
  796--799, Apr. 2018.

\bibitem{Bernier2019}
\BIBentryALTinterwordspacing
C.~Bernier, F.~Dehmas, and N.~Deparis, ``{Low Complexity { LoRa} Frame
  Synchronization for Ultra-Low Power Software-Defined Radios},'' Sep. 2019.
  [Online]. Available: \url{https://hal-cea.archives-ouvertes.fr/cea-02280910}
\BIBentrySTDinterwordspacing

\bibitem{Xhonneux2019}
M.~Xhonneux, D.~Bol, and J.~Louveaux, ``A low-complexity synchronization scheme
  for {LoRa} end nodes,'' \emph{ArXiv e-prints}, Dec. 2019,
  \url{https://arxiv.org/abs/1912.11344}.

\bibitem{Tapparel2020}
J.~{Tapparel}, O.~{Afisiadis}, P.{ Mayoraz}, A.~{Burg}, and A.{
  Balatsoukas-Stimming}, ``An open-source {LoRa} physical layer prototype on
  {GNU} radio,'' \emph{ArXiv e-prints}, Feb. 2020,
  \url{https://arxiv.org/abs/2002.08208}.

\bibitem{Abeele2017}
F.~{Van den Abeele}, J.~{Haxhibeqiri}, I.~{Moerman}, and J.~{Hoebeke},
  ``Scalability analysis of large-scale {LoRaWAN} networks in ns-3,''
  \emph{IEEE Internet of Things Journal}, vol.~4, no.~6, pp. 2186--2198, Dec.
  2017.

\bibitem{Reynders2018}
B.~Reynders, Q.~Wang, and S.~Pollin, ``A {LoRaWAN} module for {ns-3}:
  Implementation and evaluation,'' in \emph{Workshop on ns-3}, ser. WNS3
  '18.\hskip 1em plus 0.5em minus 0.4em\relax New York, NY, USA: ACM, 2018, pp.
  61--68.

\bibitem{Bor2016}
M.~C. Bor, U.~Roedig, T.~Voigt, and J.~M. Alonso, ``Do {LoRa} low-power
  wide-area networks scale?'' in \emph{ACM Int. Conf. on Modeling, Analysis and
  Simulation of Wireless and Mobile Systems}, ser. MSWiM '16.\hskip 1em plus
  0.5em minus 0.4em\relax New York, NY, USA: ACM, 2016, pp. 59--67.

\bibitem{Pop2017}
A.~{Pop}, U.~{Raza}, P.~{Kulkarni}, and M.~{Sooriyabandara}, ``Does
  bidirectional traffic do more harm than good in {LoRaWAN} based {LPWA}
  networks?'' in \emph{IEEE Global Commun. Conf. (GLOBECOM)}, Dec. 2017, pp.
  1--6.

\end{thebibliography}

\end{document}